\documentclass[12pt]{article}

\usepackage{euscript}
\usepackage{amssymb}
\usepackage{amsfonts}
\usepackage{amsbsy}
\usepackage{amsmath}
\usepackage{epsfig}
\usepackage{hyperref}

\def\IZ{\mathbb{Z}}
\def\IR{\mathbb{R}}

\def\CM {{\cal M}}
\def\CR {{\cal R}}
\def\CN {{\cal N}}

\def\re{\mbox{Re }}
\def\im{\mbox{Im }}

\def\bj{\bar{j}}

\def\bz{\bar{z}}
\def\bF{\bar{F}}

\newcommand{\be}{\begin{equation}}
\newcommand{\ee}{\end{equation}}
\newcommand{\ba}{\begin{eqnarray}}
\newcommand{\ea}{\end{eqnarray}}

\newcommand{\ft}{\footnote}

\def\a{\nu}

\numberwithin{equation}{section}

\usepackage{sectsty}
\sectionfont{\sffamily}
\subsectionfont{\sffamily}
\subsubsectionfont{\sffamily}

\begin{document}

\begin{flushright}
IC/2005/07\\
SISSA-10/2005/EP
\end{flushright}
\begin{center}
\huge{\sc Statistics of $M$ theory Vacua.}\\
\bigskip\bigskip
\large{ {\sc B.S. Acharya$^a$ \ft{bacharya@ictp.it}, F. Denef$^b$
\ft{denef@physics.rutgers.edu} and R. Valandro$^c$
\ft{valandro@sissa.it}}}
\\
\bigskip\normalsize
{\sf $^a$ Abdus Salam ICTP,
Strada Costiera 11,\\ 34014 Trieste, ITALIA\\
\smallskip
$^b$ New High Energy Theory Center,\\ Rutgers University, Piscataway.\\
NJ 08554 USA\\
\smallskip
$^c$ International School for Advanced Studies (SISSA/ISAS),\\Via Beirut 2-4, 34014 Trieste, ITALIA }

\end{center}
\bigskip
\begin{center}
{\bf {\sc Abstract}}
\end{center}

We study the vacuum statistics of ensembles of $M$ theory
compactifications on $G_2$ holonomy manifolds with fluxes, and of
ensembles of Freund-Rubin vacua. We discuss similarities and
differences between these and Type IIB flux landscapes. For the
$G_2$ ensembles, we find that large volumes are strongly suppressed,
and for both, unlike the IIB case, the distribution of cosmological
constants is non-uniform. We also argue that these ensembles
typically have exponentially more non-supersymmetric than
supersymmetric vacua, and show that supersymmetry is virtually
always broken at a high scale.

\newpage


\section{Introduction and Summary}

String theory has long held the promise to provide us with a
complete and final description of the laws of physics in our
universe. Yet from the start it was clear that the theory allows
many vacua we clearly do not live in --- ten dimensional Minkowski
space being the simplest example. Over the past two decades, the
existence of an infinite set of distinct stable vacua in which we
obviously do not reside was firmly established. For some of those,
such as Type IIB string theory on ${\rm AdS}_5 \times S^5$ with
flux, we even have complete nonperturbative descriptions. The vast
majority of these well understood vacua have unbroken supersymmetry
and come either with massless scalars or a negative cosmological
constant of the order of the KK scale, all properties we certainly
do not observe.
Moreover, even though constructing vacua with light charged particles
is fairly straightforward, typically such vacua are very different
from the standard model at low energies.

The implicit assumption, of course, was that at least one
vacuum should exist with broken supersymmetry, with just the
standard model at low energies, and a couple of rather astonishing
hierarchies between the cosmological, electroweak and Planck scales.
One could hope that some deep dynamical mechanism uniquely
selects such a vacuum, with all its peculiarities,
rather than one from the enormous set of consistent vacua already
firmly established, or from the presumably even much more enormous
set of vacua that do not satisfy the rather artificial constraint of
being under our full computational control at the present stage of
theoretical development. Thus far however, no indication whatsoever
has emerged that this is indeed the case, and anyway there is no
realistic hope that the ``correct'' vacuum could be computed from
such hypothetical dynamical principles with what we currently know
about string theory.

It thus seems more modest and productive to follow the historically
successful approach of trying to construct models that are
compatible both with experimental and theoretical constraints.
Whilst progress is continually being made in the program of
constructing string theory vacua which meet rough observational
requirements such as obtaining the standard model spectrum at low
energies, no massless scalars or a small positive cosmological
constant, no explicit model has been constructed combining all of
these features. Clearly though, the examples studied thus far
represent only an infinitesimal part of the tip of the iceberg, as
computational complexity quickly turns into a major obstacle. The
iceberg in question has been called the string theory landscape, and
the problem we face is to find places in this landscape that could
describe our universe.

Before embarking on this program however, it would evidently be
useful to know what our chances of success are. For example it could
save us a lot of time and effort if we managed to exclude large
regions of the landscape, or more precisely large classes of models,
before attempting to find realistic string theory vacua there.
Traditionally the approach has been to focus on the most easily
controllable constraints such as light charged particle spectra,
while ignoring issues such as moduli stabilization, supersymmetry
breaking or the cosmological constant. However these constraints are
equally important, and it would be very useful to estimate how much
they reduce the set of possibilities. For instance,  one could
propose a certain compactification with just the standard model
living on a set of branes, and with very large compactification
volume to explain the hierarchy between the electroweak and Planck
scales, along the lines of \cite{ADD}. This would already seem like
a considerable triumph. A courageous model builder might then go on
and try to turn on various fluxes to stabilize all moduli within the
required region of moduli space, following the scenario of
\cite{kklt} or \cite{Acharya:2002kv}. But this theorist will
plausibly end his project disillusioned, finding not a single vacuum
with volume even remotely close to what is required for this model
to work. Indeed, it is generally true in these scenarios that flux
vacua rapidly become more scarce at larger compactification volumes,
and completely cease to exist even at very modest sizes. Clearly,
our model builder would have benefited greatly from some simple
estimates of distributions of actual vacua over the parameter space
of the model.

In \cite{Bousso:2000xa}, it was furthermore pointed out that a
sufficiently fine ``discretuum'' of string theory vacua could
accommodate an extremely small but nonzero cosmological constant. To
find out whether this idea can be realized in a given ensemble of
vacua, and more generally to analyze how strongly constraints on
various parameters reduce the number of possibilities within a given
class of models (possibly down to zero as in the example above), one
would like to have an estimate of how many vacua with certain
properties lie in a given region of the landscape. In other words,
one needs to study the statistics of vacua in this region (where
``statistics'' does {\it not} refer to any probability measure, but
simply to number distributions on parameter space). This program was
initiated in \cite{Douglas:2003um} and has been developed, thus far
mainly for Type IIB flux vacua, in
\cite{Ashok:2003gk,Douglas:2004kp,Douglas:2004zu,Denef:2004ze,Giryavets:2004zr,
Douglas:2004kc,Misra:2004ky,Douglas:2004zg,Conlon:2004ds,Kumar:2004pv,DeWolfe:2004ns}.
A number of ideas on the important question of distributions of
supersymmetry breaking scales were proposed in
\cite{Susskind:2004uv,Douglas:2004qg,Dine:2004is}, and subsequently
significantly qualified in \cite{Denef:2004cf}. Further
considerations on this topic have been made from an effective field
theory point of view in \cite{Dine:2004ct,Dine:2005yq}.
Distributions of gauge groups and particle spectra were studied in
detail in \cite{Dijkstra:2004cc,Blumenhagen:2004xx}. A number of
effective field theory models of the landscape and phenomenological
consequences have been analyzed in
\cite{Dvali:2004tm,Dienes:2004pi,Tegmark:2004qd,Arkani-Hamed:2005yv},
and some aspects of the statistics of string vacua beyond critical
theories have been explored in \cite{Silverstein:2004sh}. Finally,
more critical viewpoints of the string landscape have appeared in
\cite{Banks:2003es,Banks:2004xh}.

One class of vacua that are interesting from a
phenomenological point of view are $M$ theory compactifications on
manifolds of $G_2$ holonomy \cite{g2phen}.
Because of string duality there are presumably
at least as many such vacua as there are
heterotic string vacua on Calabi-Yau threefolds.
It is thus important to get an idea of how
moduli, compactification volumes, cosmological constants and
supersymmetry breaking scales are distributed within this class of
models. This will be the subject of the main part of this paper.

Although guidance for model building already provides enough
motivation for statistical studies, one could be more ambitious and
try to infer general properties of distributions of string theory
vacua from statistical analysis of particular regions of the
landscape combined with genericity arguments. To get some confidence
in such results, one would need to explore and compare as many
regions of the landscape as possible. Thus far however, only Type
IIB flux vacua have been analyzed in detail. Because of string
duality, one might optimistically hope that such IIB flux vacua
could be representative --- in the sense that they constitute a
significant fraction of all string vacua, perhaps constrained to
have some additional properties such as supersymmetry in the UV.
However, it should be noted that the studies in \cite{Ashok:2003gk}
- \cite{DeWolfe:2004ns} are limited to vacua described as Calabi-Yau
orientifolds at moderately large volume and moderately weak string
coupling, and strictly speaking these represent only a corner of the
string theory landscape. In principle, distributions of observables
could change dramatically as one explores different regions of the
landscape. It was suggested in \cite{obs} that the set of four
dimensional string and $M$ theory vacua with $\CN=1$ or no
supersymmetry is a disconnected space whose different components
represent qualitatively different low energy physics, and this could
translate into very different statistical properties. This gives us
additional motivation to study the statistics of $G_2$ vacua and
compare to the IIB case.

Another branch of the landscape
whose statistical analysis we initiate here is the set of
Freund-Rubin vacua \cite{Freund:1980xh}, i.e.\ $M$ theory
compactifications on Einstein manifolds with positive scalar curvature. Their properties are quite
different from more familiar compactifications on special holonomy
manifolds, and as we will see, this is reflected in their vacuum distributions
which are very different as well.

Finally, one could take things one step further and altogether
discard the idea that some mysterious dynamical mechanism has
uniquely selected our vacuum and in particular picked the bizarre
and exceedingly nongeneric scale hierarchies which just happen to be
also necessary to make structure formation and atoms other than
hydrogen or helium possible
\cite{Weinberg:1987dv,Vilenkin:1994ua,Agrawal:1997gf,Arkani-Hamed:2004fb,Arkani-Hamed:2005yv}.
Instead, one could start with the hypothesis that a ``multiverse''
exists in which a huge number of vacua is actually realized, and in
particular that we observe ourselves to be in a vacuum with such
large scale hierarchies simply because this is needed for structure
and atoms, and therefore observers, to exist. To make direct
predictions from just string theory in such a framework, one would
need to know the probability measure on at least the part of the
landscape compatible with a number of basic requirements. There is
no established way of computing these probabilities at this time,
but as an additional working hypothesis one might assign for example
in a given ensemble of flux vacua approximately the same probability
to every choice of flux. One could refine this by restricting this
equal probability postulate to subsets of vacua with fixed values of
parameters relevant for cosmology, such as the vacuum energy, as one
does for microstates with equal energy in the microcanonical
ensemble of statistical mechanics. Different choices of these
relevant parameters might then be weighted by cosmological
considerations (up to the extent that this is needed, as some will
be effectively fixed by environmental requirements). Under such
hypotheses, suitable number distributions can be interpreted as
probability distributions, and one can test the hypotheses that went
in by Bayesian inference.

It should be emphasized that this framework is significantly more
predictive than the traditional model building approach of simply
considering any model compatible with current observations. Under
these simple hypotheses, the landscape picture together with a few
rough environmental principles gives a new notion of naturalness for
effective field theories, which translates into a set of rules for
model building. This turns out to lead to \emph{very distinct}
models which do not need contrived engineering to fit known data,
and which moreover give very specific predictions, including many
unambiguous signatures at LHC
\cite{Arkani-Hamed:2004fb,Arkani-Hamed:2004yi,Arkani-Hamed:2005yv}.

More theoretical data on number distributions obtained from string
theory would obviously be very useful to make further progress in
this area, and although as we discussed there are several other
motivations for studying the statistics of string and $M$ theory
vacua, we consider this to be a very important one.

The analysis of the $M$ theory vacua we study in this paper is
somewhat technical, so for the ease of the reader, in the remainder
of this introduction, after a brief review of the Type IIB flux
landscape, we present a summary of our main results.

\subsection{Review of Type IIB statistics}

For comparison, let us briefly review what is known about the
statistics of IIB flux vacua of Calabi-Yau orientifolds. There is a
natural splitting of K\"ahler and complex structure moduli in this
context. Turning on flux induces a superpotential \cite{gvw} which
only depends on the complex structure moduli. In suitable
circumstances the K\"ahler moduli can be stabilized by
nonperturbative effects, along the lines of
\cite{kklt,Denef:2004dm}. It is reasonable to ignore the K\"ahler
moduli altogether as far as vacuum statistics is concerned, because
(1) the main contribution to vacuum multiplicities comes from the
huge number of different fluxes, (2) at sufficiently large volume
the K\"ahler moduli do not influence the positions in complex
structure moduli space significantly, and (3) practically, the
K\"ahler sector of is less under control and more difficult to treat
systematically.

The results for the distributions of moduli, cosmlogical constants,
volumes and supersymmetry breaking scales for IIB flux vacua are
roughly as follows.

\begin{enumerate}

\item The number of supersymmetric vacua in a region ${\cal R}$ of complex structure
moduli space is estimated by \cite{Ashok:2003gk}
\begin{equation}
 \CN_{\rm vac}({\cal R}) = \frac{(2 \sqrt{\pi} L)^{b_3}}{b_3!} \int_{\cal
 R}  \det(R+\omega {\bf 1})
\end{equation}
where $L$ is the induced D3 brane charge by O3, O7 and D7 branes,
$b_3$ is the third Betti number of the Calabi-Yau manifold, $R$ is
the curvature form on the moduli space and $\omega$ the K\"ahler
form. Actually this expression gives an index rather than an
absolute number --- it counts vacua with signs, so it is strictly
speaking a lower bound. Essentially this expression implies that
vacua are uniformly distributed over moduli space, except when the
curvature part becomes important, which is the case near conifold
degenerations. The above result was verified by Monte Carlo
experiments in \cite{Giryavets:2004zr, Conlon:2004ds}.

\item The cosmological constant for supersymmetric vacua is $\Lambda = - 3 e^K
|W|^2$.\footnote{Here and in the following we work in string units.}
Its distribution for values much smaller than the string scale was
found in \cite{Denef:2004ze} to be essentially uniform, i.e.\
\begin{equation} \label{IIBccdistr}
 d\CN \sim  \CN_{\rm tot} \, d\Lambda.
\end{equation}
Here $\CN_{\rm tot} \sim L^{b_3}/b_3!$ is the total number of flux
vacua.

\item The compactification volume $V$ is stabilized by nonperturbative D3-instanton
effects and/or gaugino condensates, both of which give contributions
$\sim e^{-c V^{2/3}}$ to the superpotential, where $c<1$ decreases
when $b_3$ increases. In the scenario of \cite{kklt}, these have to
balance against the contribution $W_0$ from the fluxes. That is, at
sufficiently large $V$ (or equivalently sufficiently small $W_0$),
$e^{-c V^{2/3}} \sim W_0$. Since $|W_0|^2 \sim \Lambda$ is uniformly
distributed according to (\ref{IIBccdistr}), this gives the volume
distribution
\begin{equation}
 d\CN \sim \CN_{\rm tot} \, e^{-2 c V^{2/3}} d (V^{2/3}).
\end{equation}
Large volumes are therefore exponentially suppressed, and for
reasonable values of $L$ and $b_3$, the maximal volume will be of
the order $V_{\rm max} \sim \left(\frac{\log \CN_{\rm tot}}{2
c}\right)^{3/2} \sim (b_3/c)^{3/2}$.\footnote{After this paper was
sent out, an interesting paper appeared
\cite{Balasubramanian:2005zx} pointing out the possible existence of
nonsupersymmetric AdS vacua at larger volumes, through a subtle
balancing of nonperturbative and perturbative corrections in the
potential \cite{Balasubramanian:2004uy}.} One could also interpret
the IIB complex structures fixed by the fluxes to be mirror to IIA
K\"ahler moduli. Then the distribution of IIA compactification
volumes can be shown to be \cite{DDlargevol}
\begin{equation}
 d \CN \sim \frac{(k L)^{b_3}}{b_3!} \, d(V_{IIA}^{-b_3/6})
\end{equation}
for $V_{IIA}\gg 1$. Here $k$ is a constant weakly decreasing with
increasing $b_3$. Again, large volumes are suppressed, now bounded
by $V_{IIA}<(e k L/b_3)^6$.

\item The flux potential has nonsupersymmetric minima as well.
The supersymmetry breaking scale $F=e^{K/2}|DW|=M_{\rm susy}^2$ for
$F \ll 1$ is distributed as \cite{Denef:2004cf}
\begin{equation}
 d \CN \sim \CN_{\rm tot} \, dF
\end{equation}
if no further constraints are imposed, and
\begin{equation}
 d\CN \sim \CN_{\rm tot} \, F^5 dF \, d\Lambda
\end{equation}
if one requires the cosmological constant $\Lambda$ to be much
smaller than $F^2$. Scenarios in which supersymmetry breaking is
driven by D-terms were also considered in \cite{Denef:2004cf}, and
it was pointed out that in the special case of supersymmetry
breaking by an anti-D3 brane at the bottom of a conifold throat, low
scales are more natural. Since we work in the large radius regime,
there is no counterpart of this scenario in the M-theory
compactifications we will study, so we will not get into details
here. We should also point out that large classes of string
compactifications have been proposed in \cite{Saltman:2004jh} where
supersymmetry is broken at the KK scale.

\end{enumerate}

Next we summarise our results on the statistics of Freund-Rubin
vacua.

\subsection{Freund-Rubin statistics}

Freund-Rubin vacua of $M$ theory have geometry ${\rm AdS}_4 \times
X_7$, with $X_7$ a positively curved Einstein manifold, and can be
understood as arising from the near-horizon geometry of $N$
coincident M5-branes, which become $N$ units of $G_4$-flux in the
AdS-space. The compactification volume is fixed and depends on $N$
and the choice of $X_7$. Typically, these geometries cannot really
be considered as compactifications on $X_7$ in the usual sense,
because the Kaluza-Klein scale tends to be of the same order as the
AdS scale.

Nevertheless we can study the distributions of AdS cosmological
constants $\Lambda$ and compactification volumes $V$. We will do
this in section \ref{sec:FR} for a model ensemble with $X_7 =
X/\IZ_k$, where we vary $k$ and $N$. This ensemble is extremely
simple, and is therefore additionally useful as a simple toy model for
testing counting methods.

At fixed $k$ ie for a fixed topology of the extra dimensions, we find the following distributions for $\Lambda$ and
$V$:
\begin{equation}
 d \CN(\Lambda) \sim d \Lambda^{-2/3}, \qquad d \CN(V) \sim d
 V^{6/7}.
\end{equation}
This already shows a dramatic difference compared to IIB flux vacua.
First, there are an infinite number of vacua, since $N$ is arbitrary.
Secondly, the distribution of $\Lambda$ is not uniform near zero,
but diverges. Finally, large volumes are not suppressed --- the
larger $N$, the larger the volume becomes.

Allowing both $k$ and $N$ to vary ie by sampling the topology of the extra
dimensions as well, these results significantly
change. Now
\begin{equation}
 d \CN(\Lambda) \sim d \Lambda^{-2}, \qquad d \CN(V) \sim d
 V^6.
\end{equation}
The qualitative features are the same though: smaller cosmological
constants and larger volumes are favored.\footnote{Perhaps we should
stress that by ``favored'' we do not mean ``more probable''. As
emphasized earlier, we are computing \emph{number} distributions at
this level, not probability distributions.} We also obtain the joint
distribution for $V$ and $\Lambda$:
\begin{equation}
 d\CN(V,\Lambda) \sim \Theta(V,\Lambda) \, \frac{dV}{V^4} \, \frac{d\Lambda}{\Lambda^4}
\end{equation}
where $\Theta(V,\Lambda)=1$ when $V^{-3} \leq \Lambda \leq V^{-9/7}$
and zero otherwise (see also fig.\ \ref{vacpoints}). One interesting
feature that can be read off from this distribution is that at fixed
$\Lambda$, $V$ actually accumulates at \emph{smaller} values,
opposite to what we found for the unconstrained case. This is
possible because the step function $\Theta$ allows $\Lambda$ to vary
over a larger domain when $V$ increases. This illustrates the
importance of constraints for statements about which parameters are
favored. Such issues become especially important if one wishes to
interpret number distributions as probability distributions, since
through such correlations, the dependence of these probabilities on
one parameter may strongly influence the likelihood of values of the
other.

\subsection{${\bf G_2}$ holonomy statistics}

The main part of this paper is devoted to the study of $G_2$
holonomy vacua. These have $\CN=1$ supersymmetry and as usual come
with many massless moduli. In \cite{Acharya:2002kv}, a mechanism was
proposed to stabilize all geometric moduli, basically by combining
$G_4$-fluxes (which alone do not lead to any stable
vacua\footnote{This is true in the classical supergravity
approximation, in which we work throughout this paper. The results
of \cite{Kachru:2004jr} indicate that corrections to the K\"ahler
potential may change this.}) with a contribution to the potential
induced by nonabelian degrees of freedom living on a codimension
four singularity in the $G_2$ manifold. In the low energy effective
theory this has simply the effect of adding a complex constant
$c=c_1 + i c_2$ to the flux induced superpotential, given by a
complex Chern-Simons invariant on the singular locus. The physics of
this is presumably related to the Myers effect on D6-branes in type
IIA, although this remains to be clarified. Anyway, regardless of
its actual realization, this is also the simplest modification of
the flux superpotential one can imagine and which leads to stable
vacua.

Unfortunately, no explicit constructions of $G_2$ manifolds with the
required properties are known (since $G_2$ manifolds themselves
are technically difficult to make), so we will just take these ensembles
of flux superpotentials shifted with a constant as our starting
point. Another problem is that not
much is known about the metric on $G_2$ moduli spaces, apart from the
fact that they are derived from a K\"ahler potential $K = - 3 \log
V$. To get around this problem, we follow two approaches. One,
detailed in section \ref{sec:general}, is to keep the K\"ahler
potential completley arbitrary and get general results. This will allow us to
get quite far already. For some purposes however, more information
is needed. In section \ref{sec:statmodel}, we introduce an ensemble
of model K\"ahler potentials which satisfy all known constraints for
$G_2$ moduli spaces. The models turn out to be exactly solvable, in
the sense that all supersymmetric and nonupersymmetric vacua can be
found explicitly, and we push the vacuum statistics analysis to the
end.

The results we find for general ensembles are as follows.
\begin{enumerate}

\item The number of supersymmetric vacua in a region $\CR$ of moduli space
is, for large $c_2$:
\begin{equation}
 \CN_{\rm susy} \sim c_2^{b_3}\,
 \mbox{Vol}(\CR).
\end{equation}
Here $b_3$ is the third Betti number of the $G_2$ manifold, which is
the number of moduli as well as the number of fluxes. Thus, vacua
are distributed uniformly, at least in the large radius region of
moduli space, where classical geometry and our analysis are valid.
Note in particular that despite the absence of a tadpole constraint
like in IIB, the number of vacua in the region of moduli space where
our analysis is valid is finite. The result is quite similar to what
was found for IIB flux vacua, with $c_2$ playing the role of the
tadpole cutoff $L$.

\item The distribution of volumes $V$ (measured in 11d Planck units) is given by
\begin{equation}
 d\CN = (k c_2)^{b_3} d V^{-3b_3/7}.
\end{equation}
Here $k$ is a constant weakly decreasing with increasing $b_3$.
Thus, as in the IIB ensemble and its IIA mirror, large volumes are
suppressed, and strongly so if $b_3$ is large. An upper bound on $V$
is given by
\begin{equation} \label{Vbound}
 V < (k c_2)^{7/3}.
\end{equation}

\item It can be shown that in a supersymmetric vacuum
\begin{equation}
 \Lambda \sim \frac{c_2^2}{V^3},
\end{equation}
where $\Lambda$ is expressed in 4d Planck units. The distribution of
supersymmetric cosmological constants is therefore entirely
determined by the distribution of volumes:
\begin{equation}
 d\CN = (k c_2^{5/7})^{b_3} d \Lambda^{b_3/7}.
\end{equation}
In particular for large $b_3$, small cosmological constants are
strongly suppressed, and bounded below by $\Lambda > 1/c_2^5 k^7$.
This is radically different from the IIB case, where this
distribution is uniform, and cosmological constants can be obtained
as low as $1/\CN_{\rm tot}$. The underlying reason for this
difference is that in the IIB ensembles which have been studied,
there are four times more fluxes than moduli, while in our $G_2$
ensemble there are only as many fluxes as moduli. This means there
is much more discrete tunability in the IIB case, which makes it
possible to tune the cosmological constant to a very small value.

At this point the reader may wonder if this does not blatantly
contradict what one would expect from string duality. This is not
so, because our $G_2$ ensemble is not dual to the standard IIB
ensemble with both NS and RR 3-form fluxes. On both sides, some of the
flux degrees of freedom do not have a counterpart as flux degrees of
freedom on the other side. In favorable circumstances these are dual
instead to discrete geometric deformations away from special
holonomy, although this has not been established in general
\cite{dual}.
In any case, this shows that neither of the two ensembles is
``complete''. Conceivably, relaxing the $G_2$ holonomy condition to something
weaker could allow the cosmological constant in the $M$ theory
ensemble to be much more finely tuned, and one could imagine that
one would then get a uniform distribution again, although this is by
no means certain. It would be very interesting to verify this, but
at present a detailed description of such would be vacua is unknown.

\item The distribution of supersymmetry breaking scales is similarly
quite different from the IIB case, again essentially because of the
lack of discrete tunability. Because there are as many fluxes as
moduli, one can express all flux quanta as a function of the susy
breaking parameters $F_i$ at a given point in moduli space. The
equation $V'=0$ at that point thus becomes a complicated quadratic
equation in the $F_i$, whose solutions, apart from the
supersymmetric $F=0$, are not tunably small. This means the analysis
of \cite{Denef:2004cf} does not apply, and general, complete
computations of distributions become hard.

A number of simple observation can be made though. Since we have a
system of $b_3$ quadratic equations in $b_3$ variables $F_i$, the
number of nonsupersymmetric branches, for a fixed choice of flux can
be up to $2^{b_3}$ (minus the supersymmetric vacuum).
In the model ensembles of section
\ref{sec:statmodel}, we show that this number can actually be obtained.
Since essentially all of these break supersymmetry at a high scale,
this is further support for the idea that string theory has many
more $F$-breaking flux vacua with high scale breaking than with low
scale.

More precisely, we show that in these ensembles, any vacuum with
cosmological constant $\Lambda \sim 0$ (assuming it exists) has
supersymmetry breaking scale
\begin{equation}
 M^2_{\rm susy} = \frac{r c_2}{V^{3/2}} \, m_p^2,
\end{equation}
where $r \approx 0.1$. Thus, as for the cosmological constant, the
scale of susy breaking is set by the volume $V$, and similarly its
distribution will largely be determined by the distribution of $V$.
Using the volume bound (\ref{Vbound}),\footnote{which was strictly
speaking derived for supersymmetric vacua, but the distributions of
supersymmetric and nonsupersymmetric vacua over moduli space are
expected to be very similar, as we confirm in the model ensembles of
section \ref{sec:statmodel}.} this gives a lower bound on the
supersymmetry breaking scale:
\begin{equation}
 M^2_{\rm susy} > k^{-7/2} c_2^{-5/2} m_p^2.
\end{equation}
For moderate values of $c_2$ (recalling that $k$ is decreasing with
increasing $b_3$), the supersymmetry breaking scale will therefore
always be high in these ensembles.

This behavior is quite different from Type IIB and the generic
ensembles studied in \cite{Denef:2004cf}, which again can be traced
back to the fact that there is limited discrete tunability because
there are only as many fluxes as moduli. It is conceivable that
adding discrete degrees of freedom, by allowing deformations away
from special holonomy, would bring the distribution of supersymmetry
breaking scales closer to the results of \cite{Denef:2004cf}. In any
case this would not change the qualitative conclusion that higher
superymmetry breaking scales are favored.

\end{enumerate}

Finally, in section \ref{sec:statmodel} we present and study a class
of models defined by K\"ahler potentials which give a direct product
metric on moduli space. Though very rich, the semi-classical,
supergravity vacuum structure of these models can be solved exactly,
allowing us to explicitly verify our more general results. Vacua are
labeled by $(N_i,\sigma_i)$, $i=1,\ldots,b_3$, where the $N_i$ are
the fluxes and $\sigma_i=\pm 1$. Putting all $\sigma_i=+1$ gives a
supersymmetric AdS vacuum, the $2^{b_3}-1$ other choices correspond
to nonsuperymmetric vacua. As the fluxes are varied, we find that
these are uniformly distributed over the moduli space and that there
are roughly an equal number of de Sitter vs. anti de Sitter vacua.
Not all of these vacua exist within the supergravity approximation
and we analyse the conditions under which they do, finding that an
exponentially large number survive. Finally, after analysing the
stability of these vacua we found that all de Sitter vacua are
classically unstable whilst an exponentially large number of
non-supersymmetric anti de Sitter vacua are metastable. As expected
on general grounds, the supersymmetry breaking scale is typically
high.

These model ensembles are reminiscent of the effective field theory
landscapes recently considered in \cite{Arkani-Hamed:2005yv}. In
particular, at large $b_3$, the distributions we find are sharply
peaked, see e.g.\ fig.\ \ref{ccdistr} and fig.\ \ref{susybrdistr}
for examples of distributions of cosmological constants and
supersymmetry breaking scales. The cosmological constant in these
ensembles does not scan near zero; there is a cutoff at $\Lambda
\sim -c_2^{-5}$. It is conceivable though that this would change for
more complicated K\"ahler potentials which do not lead to direct
product metrics.

\section{Freund-Rubin Statistics} \label{sec:FR}

In this section we will study the statistics of Freund-Rubin vacua.
We will study a model ensemble for which the basic distributions of
cosmological constants and volumes are extremely simple compared to
say IIB flux vacua or $G_2$-holonomy vacua. After a brief review of
Freund-Rubin vacua, included for completeness, we describe the
distributions. Note that, since 4d Freund-Rubin vacua are dual to
three dimensional CFT's, we are equivalently studing the statistics
of such theories.

\subsection{Freund-Rubin basics}

Freund-Rubin vacua \cite{Freund:1980xh} are near horizon geometries
of brane metrics in which there is an AdS factor. We will focus on
the $M$2-brane case for which the Freund-Rubin metric takes the form
 \be g_{10+1} = g_4({{\rm AdS}_a}) + a^2 g_0(X) \ee
Here ${\rm AdS}_a$ is four dimensional anti de Sitter spacetime with
radius $a$ and $X$ is a compact 7-manifold, and $g_0(X)$ is an
Einstein metric on $X$ with positive scalar curvature.:
 \be R_{ij}(g_0 ) = 6 g_{0ij} \ee
Since the actual metric on the seven compact dimensions is rescaled
by $a^2$, the scalar curvature of $X$ and the four dimensional AdS
is of order $a^{-2}$.

This means that if the volume of $X$ as measured by $g_0(X)$ is
order one compared to that of a round unit 7-sphere, then the masses
of Kaluza-Klein modes are of order the gravitational mass in AdS
(i.e.\ the inverse AdS radius). In this case, there is no meaningful
low energy limit in which one can ignore the dynamics of the
Kaluza-Klein particles.

However, if the radii of $X$ in the $g_0 (X)$ metric are smaller
than one \cite{Acharya:2003ii}, there will be a mass gap between the
gravitational fluctuations and Kaluza-Klein modes. In this situation
there is a meaningful low energy limit.

Freund-Rubin vacua can be both supersymmetric or non-supersymmetric.
In fact, for every ${\cal N}= 1$ Freund-Rubin vacuum there exists a
semi-classically stable, non-super\-sym\-metric Freund-Rubin vacuum.
This is obtained by simply changing the $M$2-branes to anti
$M$2-branes, or equivalently by changing the sign of $G$. This was
known in the Kaluza-Klein supergravity literature as
``skew-whiffing'' \cite{Duff}. In fact, there are presumably many
more non-supersymmetric Freund-Rubin vacua, since the Einstein
equations with positive curvature are a well posed problem, and this
is irrespective of supersymmetry. Unfortunately, for a general, non-supersymmetric
Freund-Rubin solution, semi-classical stability of the vacuum can only be determined
on a case by case basis. The ``skew-whiffed'' solutions are always stable.

Finally, it was pointed out only recently that Freund-Rubin vacua in
four dimensions can have chiral fermions \cite{Acharya:2003ii}. The
mechanism for this --- which evades Witten's no go theorem \cite{Wittenferm}
is that the chiral fermions are wrapped membranes localised
at singularities, precisely as in the $G_2$-holonomy case studied in
\cite{AW, Acharya:2001gy}. The possibilities for realistic particle
physics phenomena in this context are little explored at the present
time and certainly deserve further investigation. Some suggestions for
the construction of more realistic models were given in \cite{Acharya:2003ii}.

\subsubsection{Quantization of $a$} \label{sec:quanta}

The Freund-Rubin metric has one parameter $a$. On the other hand,
brane metrics also have one parameter, $N$ the number of branes.
Therefore, the two parameters are related.
The relation is
 \be a^6 = {N \over V_0} \label{aNV} \ee
where $V_0$ is the number of Planck volumes of $X$ as measured by
$g_0(X)$. A Planck volume is $L_p^7$ where $L_p$ is the eleven
dimensional Planck length.

To show that (\ref{aNV}) is correct we need the formula for the
$G$-flux in the vacuum. This is
 \be
 G \sim {1 \over a} d\,{\rm Vol}({\rm AdS}_a).
 \ee
The charge or number of branes is measured by integrating the flux
$*G$ around the brane, i.e.\ over $X$:
 \be N = \int_X *\;G = a^6 \;V_0. \ee
This reproduces (\ref{aNV}). We will write physical quantities in
terms of $N$ rather than $a$.

The four dimensional Planck scale $m_p$ is related to the eleven
dimensional $M_p$ via
 \be {m_p^2 \over M_p^2} \sim a^7 V_0 \sim {N^{7 \over 6} \over V_0^{1 \over 6}} \ee
This is the volume of $X$ in 11d Planck units. The vacuum energy is
 \be \Lambda \sim -{M_p^2 \over a^2} m_p^2 \sim - {V_0^{1 \over 2} \over N^{3 \over 2}} m_p^4 \ee

\subsection{Statistics}

Obviously at the classical level $N$ is an arbitrary integer and
there are hence an {\it infinite} number of vacua. Clearly however
as $N$ tends to infinity the volume of $X$ and the radius of AdS
also tend to infinity, and the cosmological constant goes to zero. A
more sensible question is therefore how many vacua are there with
compactification volume $V_X \leq V_*$?

Since $V_X \sim {N^{7/6} \over V_0^{1/6}}$, $V_*$ is ${N_{\rm
max}^{7/6} \over V_0^{1/6}}$. So this question has the simple
answer:
\begin{equation} \label{Vdistr0}
 \CN^0_{\rm vac}(V_X \leq V_*) = N_{\rm max} \sim  V_0^{1/7} V_*^{6/7}.
\end{equation}
Similarly the number of vacua with $|\Lambda | \leq \Lambda_*$ is
infinite. However, since $|\Lambda | \sim V_0^{1/2} N^{-3/2}$ the
number of vacua with $|\Lambda | \geq \Lambda_*$ in 4d Planck units
is
\begin{equation} \label{Lambdadistr0}
 \CN^0_{\rm vac}(|\Lambda| \geq \Lambda_*) \sim V_0^{1/3}
 \Lambda_*^{-2/3}.
\end{equation}
These results are valid for a {\it fixed} Einstein manifold with
normalized volume $V_0$. We can also ask: how do these numbers
change if we vary the Einstein manifold $X$? We can construct simple
examples of ensembles of vacua in which $X$ itself is also varying
as follows.

If $X$ has a $U(1)$ symmetry then we can quotient by a discrete
subgroup ${\bf Z_k} \subset U(1)$. The quotient $X/{\bf Z_k}$ has
volume $V_0 ({X \over {\bf Z_k}}) = {V_0(X) \over k}$, but is still
Einstein with the same cosmological constant. There are many
examples of such families. A very simple one has $X$ $=$ $S^7$ the
round 7-sphere, regarded as the sphere in ${\bf C^4}$. This solves
the equations of motion. If the coordinates of ${\bf C^4}$ are
denoted by $z_i$, then ${\bf Z_k}: z_i \rightarrow e^{2\pi i \over
k} z_i $ acts freely on $S^7$. The quotients ${S^7 \over {\bf Z_k}}$
provide a family of topologically distinct Freund-Rubin solutions
labeled by $k$.


\begin{figure}
\begin{center}
  \epsfig{file=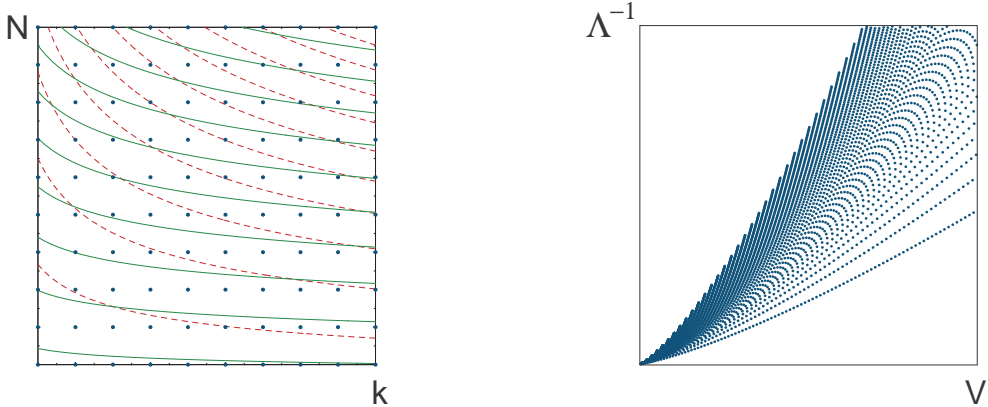,height=6.5cm,angle=0,trim=0 0 0 0}
  \caption{\footnotesize \emph{Left}: Lattice of vacua in $(k,N)$-space.
  The green solid lines have constant $V$ and the red dashed lines constant
  $\Lambda^{-1}$. Both are increasing with $N$ and $k$.
  \emph{Right}: Vacua mapped to $(V,\Lambda^{-1})$-space with $V \leq 200,\Lambda^{-1} \leq 2000$.
  The lower and upper boundaries correspond to $k=1$ resp.\ $N=1$.}
  \label{vacpoints}
\end{center}
\end{figure}

For such a set of vacua labeled by $(N,k)$ we have (from the
formulas in section \ref{sec:quanta})\footnote{We are ignoring
numerical coefficients which are irrelevant for our applications,
and will abuse notation and write $\Lambda$ instead of $|\Lambda|$
in what follows.}
 \ba
 kN^3 = \Lambda^{-2}\\
 kN^7 = V^6.
 \ea
We may now ask what the number of vacua is in a given region of
$(V,\Lambda)$-space. This amounts to counting lattice points in the
corresponding region in $(k,N)$-space, as illustrated in fig.\
\ref{vacpoints}. For sufficiently large regions, an estimate of this
number is given by the area in $(k,N)$-space. Such an estimate is in
general better for higher dimensional lattices and convex regions,
but gives a reasonable approximation in this two-dimensional case as
well, at least for sufficiently large regions, as we will
demonstrate below.

Instead of computing the area in $(k,N)$ space, one can integrate a
suitable vacuum number density over the region in
$(V,\Lambda)$-space. This density is given by
\begin{eqnarray}
 d\CN &=& dk \, dN \, \Theta(k-1) \, \Theta(N-1) \nonumber \\
 &=& d ( V^{-9/2} \Lambda^{-7/2} ) \, d ( V^{3/2} \Lambda^{1/2}) \,
 \Theta(V^{-9/2} \Lambda^{-7/2}-1) \, \Theta(V^{3/2} \Lambda^{1/2}-1)  \nonumber \\
 &=& \frac{3}{V^4 \Lambda^4} \, dV \, d\Lambda \,
 \Theta(1-V^9 \Lambda^7) \, \Theta(V^3 \Lambda -1) \label{VLdistr} \\
 &=& \frac{3 (\Lambda^{-1})^2}{V^4} \, dV \, d\Lambda^{-1} \,
 \Theta(\Lambda^{-1} - V^{9/7}) \, \Theta(V^3 - \Lambda^{-1})
\end{eqnarray}
Here $\Theta(x)$ denotes the step function, equal to $1$ if $x\geq
0$ and $0$ otherwise. The last line is included for comparison with
the discrete density in the $(V,\Lambda^{-1})$-plane shown in the
figure.

We see for example that \emph{at fixed $\Lambda$}, the density of
vacua $d\CN \sim dV/V^4$ is higher towards smaller volumes. However,
this does not mean that \emph{in total} there are more vacua at
smaller volumes, as the step functions cut out an allowed region for
$\Lambda$ that grows when $V$ increases. Going back to the original
parametrization in terms of $k$ and $N$ (which is more useful to
compare to the exact discrete distribution), we get that in the
continuum approximation, the total number of vacua with bounded
volume is given by
 \be {\cal N}_{\rm vac}(V_X \leq V_* ) = \int_1^{\infty}dN
 \int_1^{\infty}dk \, \Theta({V_*^6 \over N^7} - k)
 \ee
where the step function enforces the condition that we count $N$'s
and $k$'s such that the volume is bounded by $V_*$ (i.e.\ the area
under a solid green line in fig.\ \ref{vacpoints}). This integral
gives the answer
 \be \label{NFRboundedV}
 {\cal N}_{\rm vac}(V_X \leq V_*)
 = {V_*^6 \over 6} + 1 - {7V_*^{6/7} \over 6}
 \sim V_*^6.
 \ee
The subleading terms can be dropped since the continuum
approximation requires $V_* \gg 1$. The leading term here arises
from the small $N$ region so most of the vacua are here, as can be
seen in the figure as well. So we find that the total number of
vacua actually \emph{grows} with the volume, in contrast to the
decreasing $dV/V^4$ density at fixed $\Lambda$. Note that this
$V_*^6$ growth is also quite different from the number of vacua with
$V_X \leq V_*$ at fixed $k$, given by (\ref{Vdistr0}), which scales
as $V_*^{6/7}$. Therefore, sampling topologies of the extra
dimensions significantly changes the distribution of vacua.

These observations illustrate the obvious but important fact that
the distribution of a quantity can depend strongly on the
constraints and ensembles considered.

Another important issue is the validity of the continuum
approximation. Thanks to the simplicity of this ensemble, we can
compare our result (\ref{NFRboundedV}) with the exact answer:
\begin{equation}
 N_{\rm vac}(V_X \leq V_*)= \sum_{k,N \geq 1} \Theta ({V_*^6 \over N^7} - k)
 \approx \sum_{1 \leq N \leq V_*^{6/7}} \frac{V_*^6}{N^7}.
\end{equation}
The approximation we made here (neglecting integrality of $k_{\rm
max}$) becomes asymptotically exact in the large $V_*$ limit. In the
same limit, the sum over $N$ gives a factor $\zeta(7)$, so
\begin{equation}
 N_{\rm vac}(V_X \leq V_*) \approx \zeta(7) V_*^6.
\end{equation}
Since $\zeta(7) \approx 1.008$, this is about a factor 6 larger than
the continuum result (\ref{NFRboundedV}). The reason for this
discrepancy is the fact that the region under consideration in the
$(k,N)$-plane is very elongated along the $k$-direction, and most
contributing lattice points are at the boundary $N=1$ (see fig.\
\ref{vacpoints}), so the integral is not a very accurate
approximation of the sum. The continuum approximation will be better
for regions in $(V,\Lambda)$-space that stay away from the boundary.
Even in the case at hand though, the power of $V_*$ itself is
correct, and since the coefficient is irrelevant at the level of
accuracy we are working at anyway, the continuum approximation is
satisfactory for our purposes.

Similarly we could ask how many vacua are there with $\Lambda \geq
\Lambda_*$ (the other way around obviously gives a divergent
answer). This is
 \begin{equation}
{\cal N}_{\rm vac}(\Lambda \geq \Lambda_* ) = \int_1^{\infty}dN
\int_1^{\infty}dk \, \Theta({1 \over \Lambda_*^{2} N^3} - k) \sim
\Lambda_*^{-2}
 \end{equation}
Again this is a stronger growth than the fixed $k$ distribution
(\ref{Lambdadistr0}).

>From the joint distribution (\ref{VLdistr}), one gets the
distribution of one variable given the other one. As noted before,
at fixed $\Lambda$, vacua accumulate towards the \emph{lower} bound
on the volume $V$, which is opposite to what one has for the
distribution without constraint on $\Lambda$. For instance, when
$\Lambda_*^{-2} \leq V_*^{6}$ and $\Lambda_*^{-2/3} \geq V_*^{6/7}$,
one finds that ${\cal N}_{vac} (V\geq V_*;\Lambda \geq \Lambda_* )
\sim \Lambda_*^{-3}V_*^{-3}$ which favors both small volume and
cosmological constant.

%

{\sf In summary}: we have shown that sampling the topology of the
extra dimensions drastically changes the distribution of
cosmological constants and volumes (i.e.\ eleven dimensional Planck
scales). A priori, Freund-Rubin vacua statistically congregate in
large volume, small cosmological constant regions. However, we saw
that strong correlations between cosmological constant and volume
exist, causing the effective distribution of one quantity to depend
significantly on the constraints imposed on the other. For example
at fixed cosmological constant, vacua actually prefer the lower
volumes within the allowed range. Finally, we stressed
that there exist at least as many non-supersymmetric, meta-stable Freund-Rubin
vacua than supersymmetric.

One important point to keep in mind is that we have sampled a very
simple set of Einstein manifolds. One would certainly like to know
how representative the distribution of volumes and vacuum energies
for the $(N,k)$ ensembles we studied is in the ensemble of all
Einstein manifolds with positive scalar curvature. Some more
complicated ensembles of Einstein 7-manifolds are described in
\cite{Einstein1, Einstein2} and \cite{Einstein1} gives a formula for
the volumes which could be used to study the distributions of vacua
as we have done here.

\section{$\mathbf{G_2}$ Holonomy Statistics I: general results}
\label{sec:general}

$G_2$-holonomy vacua are compactifications of $M$ theory to four
dimensions which, in the absence of flux classically give 4d ${\cal
N}=1$ vacua with zero cosmological constant. These classical vacua
have $b_3(X)$ complex moduli, of which the real parts are axions
$t_i$ and the other half $s_i$ are the massless fluctuations of the
metric on $X$.

The addition of fluxes when $X$ is smooth does {\it not} stabilise
these moduli, as the induced potential is positive definite and runs
down to zero at infinite volume. However, if $X$ has an orbifold singularity
along a 3-manifold $Q$, additional non-Abelian degrees of freedom
arise from massless membranes \cite{bsaYM}. Nonabelian flux for
these degrees of freedom then gives an additional contribution to
the potential which {\it can} stabilise all the moduli if $Q$ admits
a complex, non-real Chern-Simons invariant \cite{Acharya:2002kv}.
This is the case if, for instance, $Q$ is a hyperbolic manifold.

The vacua studied in \cite{Acharya:2002kv} were supersymmetric with
negative cosmological constant. In fact, it was shown that in the
large radius approximation, for a given flux within a certain range,
there is a single supersymmetric vacuum (in addition to an unstable
de Sitter vacuum). In principle however there could be other,
non-supersymmetric, vacua and one of our aims here is to study this
possibility. One might wonder if any of these vacua could be
metastable de Sitter. We will answer this question to a certain
extent.

One of the difficulties in studying $G_2$-holonomy compactifications
is that $G_2$-holonomy manifolds are technically very difficult to
produce. For instance, we still do not know whether or not there
exists a $G_2$-holonomy manifold with a non-real Chern-Simons
invariant. So how can we hope to study the statistics of such vacua?
As we will see below the superpotential of these
$G_2$-compactifications with flux is very simple and does not
contain much information about $X$. Instead, this information comes
through the K\"ahler potential on moduli space, which could be a
quite complicated function in general, of which very little is
concretely known. One approach then is to try to obtain general
results for an arbitrary K\"ahler potential. We will do this in the
next section by extending some of the general techniques which were
developed in the context mainly of IIB flux vacua in
\cite{Ashok:2003gk,Denef:2004ze,Denef:2004cf}. Secondly we could
study particular ensembles of model K\"ahler potentials and hope
that the results are representative in general. We will follow this
approach in section \ref{sec:statmodel}, where we study a particular
class of model K\"ahler potentials which allow explicit construction
of all supersymmetric and nonsupersymmetric vacua.

\subsection{$\mathbf{G_2}$ compactifications with fluxes and Chern-Simons invariants}

\noindent Let $X$ be the $G_2$ holonomy compactification manifold.
The complexified moduli space $\CM$ of $X$ has dimension $n=b^3(X)$
and has holomorphic coordinates $z^i$, defined by
\begin{equation}
 z^i = t^i + i s^i = \frac{1}{\ell_M^3} \int C + i \varphi,
\end{equation}
where $\varphi$ is the $G_2$-invariant 3-form on $X$ and $\ell_M$ is
given by ${1 / 2 \kappa_{11}^2} = {2 \pi / \ell_M^9}$.\footnote{In
terms of $\ell_M$, we have $T_2 = {2 \pi / \ell_M^3}$, $T_5 = {2 \pi
/ \ell_M^6}$, and $\int G / \ell_M^3 \in \IZ$. Thus, the instanton
action is $e^{2 \pi i z}$ and $t$ has periodicity 1.} The metric on
$\CM$ is K\"ahler, derived from the K\"ahler potential\footnote{Our
normalization conventions for $z$, $W$ and $K$ are slightly
different from \cite{Beasley:2002db,Acharya:2000ps}. The value of
the normalization coefficient appearing in (\ref{Kpot}) in front of
$V_X$ is verified in Appendix A.} \cite{Beasley:2002db}
\begin{equation} \label{Kpot}
 K(z,\bz) = - 3 \ln \left( 4 \pi^{1/3} V_X(s) \right),
\end{equation}
where $V_X \equiv \mbox{Vol}(X)/\ell_M^7$ is a homogeneous function
of the $s^i$ of degree $7/3$. This classical metric
will receive quantum corrections, but at large enough volumes such
corrections can be argued to be small.

We turn on 4-form flux
\begin{equation}
 G / \ell_M^3 = N_i \rho^i,
\end{equation}
where $N_i \in \IZ$ and $\rho^i$ is a basis of $H^4(X,\IZ)$, and
also assume the presence of a complex Chern-Simons contribution as
described above. This induces a superpotential
\cite{Gukov:1999gr,Acharya:2000ps,Beasley:2002db,Acharya:2002kv}
\begin{equation}
 W(z) = {1 \over \kappa_4^3} (N_i z^i + c_1 + i c_2)
\end{equation}
on $\CM$, where $c_1$ and $c_2$ are the real and imaginary parts of
the Chern-Simons invariant.

The corresponding potential is obtained from the standard four
dimensional supergravity expression (for dimensionless scalars):
\begin{equation} \label{potential}
 V = \kappa_4^2 e^K \left( g^{i\bj} F_i \bF_{\bj} - 3|W|^2 \right),
\end{equation}
where
\begin{equation}
 F_i \equiv D_{z^i} W \equiv (\partial_{z^i} + \partial_{z^i} K) W = N_i
 + \frac{1}{2i} \partial_{s^i} K \, W.
\end{equation}
We will put $\kappa_4=1$ in what follows.

Since we are working in the large radius regime, the axions $t^i$
essentially decouple from the moduli $s^i$, and all nontrivial
structure resides in the latter sector. This is seen as follows.
Writing
\begin{equation}
 W \equiv W_1(t) + i \, W_2(s), \qquad K_i \equiv \partial_{s^i} K(s),
 \qquad K_{ij} \equiv \partial_{s^i} \partial_{s^j} K = 4 g_{i \bj}
\end{equation}
and so on, the potential (\ref{potential}) becomes
\begin{eqnarray}
 V &=& e^K \left( 4 K^{ij} (\re F_i) (\re F_j) +  K^{ij} K_i K_j W_1^2 - 3 W_1^2 -
 3 W_2^2 \right) \nonumber \\
 &=& e^K \left(4 K^{ij} (\re F_i) (\re F_j) - 3 W_2(s)^2  + 4 W_1(t)^2 \right).  \label{Vsplit}
\end{eqnarray}
In the last line we used the fact that the volume $V_X$ is
homogeneous of degree 7/3, which implies the following identities:
\begin{equation} \label{Kid}
 K_i s^i = -7, \qquad K_{ij} s^j = - K_i.
\end{equation}
The second is obtained from the first by differentiation.

Since $W_1 = N_i t^i + c_1$ and everything else in (\ref{Vsplit})
depends only on $s$, it is clear that any critical point of $V$ will
fix
\begin{equation}
 N^i t_i + c_1 = 0
\end{equation}
and therefore $W_1 = \im F =0$. Apart from this, the $t^i$ are left
undetermined, and they decouple from the $s^i$. From now on we will
work on this slice of moduli space, so we can write
\begin{equation} \label{Vreal}
 V = e^K \left( 4 K^{ij} F_i F_j  - 3 W_2^2 \right)
\end{equation}
with
\begin{equation} \label{Fdef}
 F_i = D_{s^i} W_2 \equiv (\partial_{s^i} + \frac{1}{2} K_i) W_2 = N_i + \frac{1}{2} K_i (N_j s^j + c_2).
\end{equation}
The geometry of the real moduli space $\CM$ parametrized by the
$s^i$ is the real analog of K\"ahler, often called Hessian, with
metric $K_{ij}$.

\subsection{Distribution of supersymmetric vacua over moduli space}
\label{sec:susydistr}

The aim of this section is to find the distribution of vacua over
$\CM$, along the lines of
\cite{Ashok:2003gk,Denef:2004ze,Denef:2004cf}. The condition for a
supersymmetric vacuum in the above notations is
\begin{equation} \label{susycrit}
 D_{s^i} W_2(s) = 0
\end{equation}
In what follows we will drop the index $2$ to avoid cluttering. The
number of solutions in a region $\CR$ of $\CM$, for all possible
fluxes $G$ but at fixed $c_2$, is given by
\begin{equation}
 N_{\rm susy}(c_2,\CR)= \sum_{N \in \IZ^n} \int_{\CR} d^n s \, \delta^n(D_i W) |\det
 \partial_i D_j W|.
\end{equation}
The determinant factor ensures every zero of the delta-function
argument is counted with weight 1. An approximate expression
$\CN_{\rm susy}$ for the exact number of vacua $N_{\rm susy}$ is
obtained by replacing the discrete sum over $N$ by a continuous
integral, which is a good approximation if the number of
contributing lattice points is large (which will be the case for
sufficiently large $c_2$). Thus
\begin{equation} \label{approxint}
 \CN_{\rm susy} = \int_{\CR} d^n s \int d^n N \, \delta^n(D_i W) |\det
 \partial_i D_j W|.
\end{equation}
Differentiating (\ref{Fdef}) and using (\ref{susycrit}) and
(\ref{Kid}), one gets
\begin{equation}
 \partial_i D_j W = \biggl( \frac{1}{2} K_{ij} - \frac{1}{4} K_i K_j \biggr)
 W = \frac{1}{2} K_{ik} \biggl( \delta^k_j + \frac{1}{2} s^k K_j
 \biggr) W
\end{equation}
One eigenvector of the matrix in brackets is $s^j$, with eigenvalue
$-5/2$ (this follows again from (\ref{Kid})). On the orthogonal
complement $\{ v | K_j v^j = 0 \}$, the matrix is just the identity,
so all other eigenvalues are $1$. Thus,
\begin{equation}
 |\det \partial_i D_j W| = \frac{5 |W|^n}{2^{n+1}} \det K_{ij}
\end{equation}
Furthermore, by contracting (\ref{Fdef}) with $s^i$, we get at
$F_i=0$:
\begin{equation} \label{susyWval}
 W = -\frac{2}{5} c_2.
\end{equation}
To compute (\ref{approxint}), we change variables from $N_i$ to
$F_i$. The Jacobian is
\begin{equation} \label{jac}
 {\rm Jac} =|\det \partial_{N_i} F_j|^{-1} = |\det (\delta^i_j + \frac{1}{2} K_j
 s^i)|^{-1} = 2/5.
\end{equation}
Putting everything together, we get
\begin{eqnarray}
 \CN_{\rm susy} &=& \int_{\CR} d^n s \int d^n F \, \delta^n(F) \,
 \biggl(\frac{c_2}{5}\biggr)^n \det K_{ij} \\
 &=& \biggl(\frac{c_2}{5}\biggr)^n \int_{\CR} d^n s \det K \\
 &=& \biggl(\frac{4 c_2}{5}\biggr)^n \mbox{Vol}(\widehat{\CR})
 \label{susycount}
\end{eqnarray}
where $\widehat{\CR}$ is the part of the complexified moduli space
projecting onto $\CR$ (i.e.\ the direct product of $\CR$ with the
$n$-torus $[0,1]^n$ swept out by the axions $t^i$), and the volume
is measured using the K\"ahler metric $g_{i\bj} = K_{ij}/4$ on this
space.

Thus, supersymmetric vacua are distributed \emph{uniformly} over
moduli space. This result is similar to the Type IIB orientifold
case studied in \cite{Denef:2004ze}, but simpler: the Type IIB
vacuum number density involves curvature terms as well, and in order
to get a closed form expression for $n>1$, it was necessary there to
count vacua with signs rather than their absolute number.

Note that in any finite region of moduli space, the number of vacua
is finite, because $c_2$ is finite despite the absence of a tadpole cutoff on
the fluxes. In particular, the total number of vacua in the large
radius region of moduli space (where our computation is valid) is
finite. For IIB vacua on the other hand, finiteness is only obtained
after imposing the tadpole cutoff $\int F \wedge H \leq L_*=\chi/24$
on the fluxes. In a way, the Chern-Simons invariant $c_2$ plays the
role of $L_*$ here.

\vspace{0.5cm} \noindent \emph{Example} \vspace{0.3cm}

\noindent The simplest example is the case $n=1$. Then homogeneity
fixes $V_X \sim s^{7/3}$, so $K_{ss}=\frac{7}{s^2}$, and
\begin{equation} \label{nisone}
 \CN_{\rm susy}(c_2,s \geq s_*) = \frac{c_2}{5} \int_{s_*}^\infty ds
 \, \frac{7}{s^2} = \frac{7}{5} \frac{c_2}{s_*}.
\end{equation}
Thus, vacua become denser towards smaller volume of $X$, and from
this equation one would estimate there are no vacua with $s>7
c_2/5$, which is where $\CN_{\rm susy}$ drops below 1. Indeed, it is
easily verified that the exact critical point solution is $s=-7c_2/5
N$, so the largest possible value of $s$, obtained at $N=-1$, is
precisely $7 c_2/5$. Better even, from the explicit solution it
easily follows that the approximate distribution (\ref{nisone})
becomes in fact exact by rounding off the right hand side to the
nearest smaller integer.

\subsection{Large volume suppression} \label{sec:largevolsuppr}

Although the precise form of the metric for $n>1$ is unfortunately
not known, one general feature is easy to deduce: flux vacua with
large compactification volume are suppressed, and strongly so if the
number of moduli is large.\footnote{This observation was originally
made in collaboration with M.~Douglas in the context of (mirror) IIB
flux vacua \cite{DDlargevol}.} This follows from simple scaling. If
$s \to \lambda s$, $K$ shifts with a constant, so $K_{ij} \to
\lambda^{-2} K_{ij}$ and the measure $d^n s \det K \to \lambda^{-n}
\, d^n s \det K$. Thus we have for example
\begin{equation}
 \CN_{\rm susy}(c_2,s^i \geq s_*) = (c_2/s_*)^n \, \CN_{\rm susy}(1,s^i \geq 1),
\end{equation}
and
\begin{equation} \label{largevolsuppr}
 \CN_{\rm susy}(c_2,V_X \geq V_*) = k_n \left(c_2/V_*^{3/7}\right)^n,
\end{equation}
where $k_n$ is independent of $c_2$ and $V_*$.\footnote{Even though
the obvious metric divergence at $V_X = 0$ is avoided by bounding
$V_X$ from below, it might still be possible that this bound alone
does not determine a finite volume region in $s$-space. Then the
left hand side will be infinite and the scaling becomes meaningless.
In this case, additional cutoffs should be imposed, which will
complicate the dependence on $V_*$, but large volume suppression is
still to be expected.} Hence large compactification volume is
strongly suppressed when $n$ is large.

To get an estimate for the absolute number of vacua with $V_X \geq
V_*$, one would need to estimate $k_n$ in (\ref{largevolsuppr}).
This would be quite hard in general even if the metric was
explicitly known, so to get an idea let us do this for a simple toy
model, taking the volume to be the following homogeneous function of
degree $7/3$:
\begin{equation}
 V_X(s) = \biggl( \sum_i s_i^2 \biggr)^{7/6}.
\end{equation}
Then $K=-\frac{7}{2} \log(s^2)$,
\begin{equation}
 K_{ij} = \frac{7}{s^2} \biggl( 2 \frac{s_i s_j}{s^2} - \delta_{ij}
 \biggr),
\end{equation}
and
\begin{equation}
 |\det K| = \biggl( \frac{7}{s^2} \biggr)^n.
\end{equation}
This metric is actually not positive definite for $n>1$ and
therefore unphysical, but let us proceed anyway. A more sensible
model will be presented in the next section. We have $V_X \geq V_*$
iff $|s| \geq s_*$, with $s_* \equiv V_*^{3/7}$. Thus
\begin{eqnarray}
 \CN_{\rm susy}(c_2,V_X \geq V_*) &=& \biggl( \frac{7 c_2}{5} \biggr)^n
 \int_{|s|\geq s_*} d^n s \, \frac{1}{s^{2n}} \\
 &=& \biggl( \frac{7 c_2}{5} \biggr)^n \Omega_{n-1} \frac{1}{n
 s_*^n},
\end{eqnarray}
where $\Omega_{n-1} = \frac{2 \pi^{n/2}}{\Gamma(n/2)}$ is the area
of the $(n-1)$-dimensional sphere of radius 1. So
\begin{eqnarray}
 \CN_{\rm susy}(c_2,V_X \geq V_*) &=& \biggl( \frac{7 c_2}{5} \biggr)^n
 \frac{\pi^{n/2}}{(n/2)!} \frac{1}{s_*^n} \\
 &\approx& \biggl( \frac{7 \sqrt{2 \pi e}}{5} \frac{c_2}{s_* \sqrt{n}}
 \biggr)^n \approx \biggl( \frac{6 \, c_2}{\sqrt{n} s_*} \biggr)^n.
\end{eqnarray}
In going from the first to the second line we used Stirling's
approximation $m! \approx (m/e)^m$, valid for large $m$.\footnote{In
fact $m! > (m/e)^m$, so the above approximation for $\CN_{\rm susy}$
gives an upper bound.}

This confirms the general result based on scaling, but now we also
have information about the absolute numbers. In particular, this
formula suggests all vacua for this toy model must have
compactification volume approximately bounded by
\begin{equation} \label{volbound}
 V_X < \biggl( \frac{6 \, c_2}{\sqrt{n}} \biggr)^{7/3}.
\end{equation}
Similar bounds can be expected for other models, as we confirm in
section \ref{sec:statmodel}. In general, large extra dimension
scenarios with, say, micrometer scale compactification radii are
therefore excluded in these ensembles unless $c_2$ is exceedingly
large.

\subsection{Distribution of supersymmetric cosmological constants}

The vacuum energy in a supersymmetric vacuum is, in 4d Planck units,
using (\ref{susyWval}):
\begin{equation} \label{susycc}
 \Lambda = -3 e^K |W|^2 = - \alpha {c_2^2 \over V_X^3}
\end{equation}
where $\alpha = 3 / 400 \, \pi \approx 0.002$. Therefore, the only
way to get a small cosmological constant is to have large
compactification volume. This contrasts with the ensemble of type
IIB flux vacua, where very small cosmological constants can be
obtained at arbitrary complex structure (i.e.\ arbitrary mirror IIA
volume). The underlying reason is the fact that in IIB, there are
four times as many fluxes as equations $D_i W=0$, so at a given
point, there is still a whole space of (real) fluxes solving the
equations. This freedom can be used to tune the cosmological
constant to a small value. Here on the other hand, there is only one
flux per equation, so at a given $s$, the fluxes are completely
fixed, and no freedom remains to tune the cosmological constant.
This would likely change however if more discrete data were turned
on, such as the $M$ theory duals of IIA RR 2-form flux or IIB NS
3-form flux. Unfortunately these are difficult to describe
systematically in $M$ theory in a way suitable for statistical
analysis.


Let us compute the distribution of cosmological constants more
precisely. Equation (\ref{susycc}) implies that this follows
directly from the distribution of volumes. Using
(\ref{largevolsuppr}), we get
\begin{equation}
 \CN_{\rm susy}(c_2,|\Lambda| \leq \lambda_*) = k_n \biggl(
 {c_2^5 \lambda_* \over \alpha}
 \biggr)^{n/7}.
\end{equation}
The corresponding distribution density is therefore
\begin{equation}
 d\CN/d\lambda \sim \lambda^{(n-7)/7}.
\end{equation}
In particular, for $n<7$, the distribution diverges at $\lambda=0$,
while for $n>7$, the density goes to zero. In the toy model, from
(\ref{volbound}), we get an expected bound:
\begin{equation}
 |\Lambda| > {\alpha \, n^{7/2} \over 6^7 c_2^5}.
\end{equation}
For large $n$, this is much larger than the naive $1/\CN_{\rm susy}$
which was found to be a good estimate in the Type IIB case, where
the cosmological constants of supersymmetric vacua are always
distributed uniformly near zero \cite{Denef:2004ze}.

Because $G_2$ manifolds with many moduli are much more numerous than
those with only a few, we can thus conclude that small cosmological
constants are (without further constraints) strongly suppressed in
the ensemble of all supersymmetric $G_2$ flux vacua.

\subsection{Nonsupersymmetric vacua}

A vacuum satisfies $V'=0$, and metastability requires $V''>0$. Thus,
the number of all metastable vacua in a region $\CR$ is given by
\begin{equation} \label{Nvacnonsusy}
 N_{vac} = \sum_N \int_{\CR} d^n s \, \delta(V') \, |\det V''| \,
 \Theta(V'').
\end{equation}
In principle one could again approximate the sum over $N$ by an
integral and try to solve the integral by changing to appropriate
variables, as we did for the supersymmetric case in section
\ref{sec:general}, and as was done in \cite{Denef:2004cf} for
supersymmetry breaking scales well below the fundamental scale. In
practice, we will encounter some difficulties doing this for $G_2$
vacua.

By differentiating equation (\ref{Vreal}), one gets
\begin{equation}
 \partial_i V = e^K(8 (D_i D_j W) D^j W - 6 \, W D_i W)
\end{equation}
where $D_i$ denotes the Hessian Levi-Civita plus K\"ahler covariant
derivative. The matrix $M_{ij} \equiv D_i D_j W$ is related to the
fermionic mass matrix, and with this notation the critical point
condition becomes
\begin{equation} \label{nonsusycrit}
 M_{ij} F^j = \frac{3}{4} W F_i.
\end{equation}
In \cite{Denef:2004cf}, a similar equation was interpreted at a
given point in moduli space as a linear eigenvalue equation for the
supersymmetry breaking parameters $F$, assuming the matrix $M$
(denoted $Z$ there) to be independent of $F$. This is indeed the
case for the Type IIB flux ensemble, but it is not true for the
$G_2$ ensemble. The reason is again the fact that there are only as
many fluxes as moduli here, so a complete set of variables
parametrizing the fluxes $N_i$ is already given by the $F_i$
(affinely related to the $N_i$ as expressed in (\ref{Fdef})).
Therefore, all other quantities such as $W$ and $M_{ij}$ and so on
must be determined by the $F_i$. Indeed, a short computation gives:
\begin{eqnarray}
 W &=& -\frac{2}{5} (s^i F_i + c_2) \label{WF} \\
 M_{ij} &=& \frac{1}{2} \bigl(
 (K_{ij} - \frac{1}{2} K_i K_j) W + K_i F_j + K_j F_i - K^k_{ij} F_k
 \bigr) \label{MF}
\end{eqnarray}
This means that at a given point in moduli space,
(\ref{nonsusycrit}) is actually a complicated system of quadratic
equations in $F$, with in general only one obvious solution, the
supersymmetric one, $F=0$. In total one can expect up to $2^n$
solutions for $F$. The nonsupersymmetric solutions will generically
be of order $c_2$.\footnote{The physical supersymmetry breaking
scale has an additional factor $e^{K/2}$.} In particular it is not
possible to tune fluxes to make $F$ parametrically small, so it is
not possible to use the analysis of \cite{Denef:2004cf} here, and
there is no obvious perturbation scheme to compute
(\ref{Nvacnonsusy}).

\vspace{0.5cm} \noindent \emph{Example} \vspace{0.3cm}

As a simple example we take the case $n=1$. As noted in section
\ref{sec:susydistr}, homogeneity then forces $K=-7 \ln s$, hence
$M_{ss} = (7 c_2 - 5 F s)/2s^2$, and (\ref{nonsusycrit}) (considered
as an equation for $F$ at fixed $s$) has solutions $F = 0$ and
$F=\frac{14 \, c_2}{s}$. Neglecting the metastability condition, the
continuum approximated number density of nonsupersymmetric vacua is
given by
\begin{eqnarray}
 d\CN_{\rm nonsusy} &=& \int dN \, \delta(F-\frac{14\,c_2}{s}) \,
 |\partial_s (F - \frac{14\,c_2}{s})| \, ds \nonumber \\
 &=& \int \frac{2}{5} \, dF \, \delta(F-\frac{14\,c_2}{s}) \, \frac{35 \,
 c_2}{2 \, s^2} \, ds \nonumber \\
 &=& \frac{7 \, c_2}{s^2} \, ds.
\end{eqnarray}
We changed variables from $N$ to $F$ in the integral using the
Jacobian (\ref{jac}). This result should be compared to the
supersymmetric density implied by (\ref{nisone}):
\begin{equation} \label{nonsusyvsssuy}
 d\CN_{\rm nonsusy} = 5 \, d \CN_{\rm susy}.
\end{equation}
This does \emph{not} mean that for a given flux, there are on
average five nonsupersymmetric critical points for every
supersymmetric one. In fact, as we will see below, and as was
already pointed out in \cite{Acharya:2002kv}, for $n=1$, there is
exactly one nonsupersymmetric critical point for every
supersymmetric one: the supersymmetric minimum is separated from
$s=\infty$ by a barrier, whose maximum is the (de Sitter, unstable)
nonupersymmetric critical point. Equation (\ref{nonsusyvsssuy}) only
expresses that \emph{in a given region} of moduli space (at large
$s^i$), there are on average 5 times more nonsupersymmetric vacua
then supersymmetric ones. This is simply because in the region of
moduli space under consideration, the nonsupersymmetric critical
points are located at five times the value of $s$ of the
supersymmetric critical points.

\subsection{Supersymmetry breaking scales}
\label{sec:gensusybreakingscales}

For more moduli, things become more complicated. A few useful
general observations can be made though. If we require $V = 0$
(which remains a good approximation for what follows as long as $|V|
\ll e^K c_2^2$) we can make a fairly strong statement about the
value of the supersymmetry breaking scale. Contracting
(\ref{nonsusycrit}) with $s^i$ and using (\ref{WF})-(\ref{MF}) and
(\ref{Kid}), we get
\begin{equation}
 25 \, F^2 - 3 (s\cdot F)^2 - 8 \, c_2 (s \cdot F) = 0.
\end{equation}
Together with $V \sim 4 F^2 - 3W^2 =  0$, this gives a system of two
equations in two variables, $s \cdot F$ and $F^2$, with solutions:
\begin{equation}
 F^2 = \frac{3 c_2^2}{4}, \qquad s \cdot F = \frac{3 c_2}{2}.
\end{equation}
The physical supersymmetry breaking scale for such vacua (assuming
they exist) is
\begin{equation} \label{susyscalepred}
 M_{\rm susy}^2 \equiv m_p^2 \sqrt{4 e^K K^{ij} F_i F_j} = \frac{\sqrt{3} \, c_2 \, m_p^2}{8 \sqrt{\pi} \, V_X^{3/2}}
 \sim \frac{M_p^3 c_2}{m_p},
\end{equation}
where $m_p \equiv 1/\kappa_4$ and $M_p$ are the 4- and
11-dimensional Planck scales.

Let us plug in some numbers to get an idea of the implications. If
we identify $M_p$ with the unification scale $M_{\rm unif} \sim
10^{16} \, {\rm Gev}$, this means that $M_{\rm susy} \sim \sqrt{c_2}
\, 10^{14.5} \, {\rm GeV}$, and the gravitino mass $M_g \sim M_{\rm
susy}^2/m_p \sim c_2 \, 10^{10} \, {\rm GeV}$. Since $c_2$ is at
least of order 1, this estimate implies (under the given
assumptions) that in this ensemble supersymmetry is always broken at
a scale much higher than what would be required to get the
electroweak scale $M_{\rm ew} \sim 100 \, {\rm GeV}$ without fine
tuning.

In fact a slight extension of this calculation shows that even if
one allows the addition of an arbitrary constant to the potential to
reach $V=0$ (e.g.\ to model D-terms or contributions from loop
corrections), the supersymmetry breaking scale in this ensemble is
still bounded from below by the scale $c_2 M_p^3/m_p$.

One could of course question the identification $M_p \sim M_{\rm
unif}$. Lowering the 11d Planck scale down to $M_p \sim 10^{13} \,
{\rm Gev}$ for example (which requires $V_X \sim 10^{12}$) gives a
supersymmetry breaking scale $M_{\rm susy} \sim \sqrt{c_2} \,
10^{10} \, {\rm Gev}$ and gravitino mass $m_g \sim c_2 \, 10 \, {\rm
Gev}$, which for $c_2$ not too large would give low energy
supersymmetry.

Note however that in analogy with the supersymmetric case, and based
on general considerations, we expect suppression of vacua at large
volume and therefore also suppression of low energy supersymmetry
breaking scales in this ensemble. This is further confirmed by the
exactly solvable models we will present in section
\ref{sec:statmodel}. In particular we get a lower bound on the
supersymmetry breaking scale from the upper bound on the volume
$V_X$, which in general we expect to be of the form
(\ref{volbound}), i.e.\ $V_X < (c_2/r_n)^{7/3}$, with $r_n$ weakly
growing with $n$. Using the relation (\ref{susyscalepred}) between
$M_{\rm susy}$ and $V_X$, this implies the following lower bound for
the supersymmetry breaking scale
\begin{equation} \label{susybreakingbound}
 M_{\rm susy}/m_p > \frac{r_n^{7/4}}{c_2^{5/4}}.
\end{equation}
Getting $M_{\rm susy}$ below $10^{12} \, {\rm Gev}$ in the case of
many moduli would thus require $c_2$ to be at least of order $10^6$.

Using the volume distribution (\ref{largevolsuppr}), we furthermore
get an estimate for the distribution of supersymmetry breaking
scales (in 4d Planck units):
\begin{equation} \label{Fdistr}
 d\CN_{\rm nonsusy} \sim \tilde{k}_n \, c_2^{5n/7} d(M_{\rm susy}^{4n/7})/m_p^{4n/7},
\end{equation}
where $\tilde{k}_n$ is a constant independent of $c_2$ and $M_*$.

Here we have not yet taken into account the tuning required to get a
tiny cosmological constant: $|\Lambda| \sim |4F^2-3W^2|/V_X^3<
\Lambda_*$. For a given volume $V_X$ (or equivalently a given
supersymmetry breaking scale), this requires tuning the fluxes such
that $|4F^2-3W^2|<\Lambda_* V_X^3$, which can be expected to at
least add another suppression factor $\Lambda_* V_X^3 \sim
\Lambda_*/M_{\rm susy}^4$. The suppression may in fact be stronger,
if the distribution of cosmological constants is not uniform but
more like a Gaussian sharply peaked away from $\Lambda=0$. Such
distributions are quite plausible in these ensembles, as will be
illustrated by the model ensemble we will study in section
\ref{sec:statmodel}. Another potentially important factor which we
are neglecting in this analysis is the metastability constraint
(this was found in \cite{Denef:2004cf} to add another factor $M_{\rm
susy}^4$ to the distribution in the ensembles studied there).


Finally, when one also takes into account the observed value of the
electroweak scale $M_{\rm ew}$, there is an additional expected
tuning factor presumably of order $M_{\rm ew}^2 m_p^2 / M_{\rm
susy}^4$ (in the region of parameter space where this is less than
1) \cite{Susskind:2004uv,Douglas:2004qg,Dine:2004is}. Putting
everything together, this gives (for $m_p^2 > M_{\rm susy}^2 >
M_{\rm ew} m_p$):
\begin{equation}
 d \, \CN \sim \tilde{k}_n \, \Lambda_* \, M_{\rm ew}^2 \, c_2^{5n/7}
|d \, M_{\rm susy}^{4(n/7-2)}|/m_p^{4n/7-2}
\end{equation}
So we see that for $n<14$, the Higgs mass and cosmological constant
tunings tilt the balance to lower scales, while for $n>14$ higher
scales are favored, and strongly so if $n$ is large. For $n<14$ we
should keep in mind however that there is an absolute lower bound on
the supersymmetry breaking scale, given by
(\ref{susybreakingbound}), which will further be increased by the
additional tunings of cosmological constant and Higgs mass. We
should also not forget that this is only a naive analysis; in
principle a full computation of the measure should be done along the
lines of \cite{Denef:2004cf}, but as we discussed, this does not
seem possible in the present context, because of the absence of a
small parameter.

Nevertheless, the above consideration indicate clearly that low
energy supersymmetry is typically disfavored in $G_2$ flux
ensembles, and even excluded if $c_2$ is less than $10^6$.

\section{$\mathbf{G_2}$ statistics II: model K\"ahler potentials
and exact solutions} \label{sec:statmodel}

In the previous section we obtained a number of general results
about distributions of $G_2$ flux vacua, independent of the actual
form of the K\"ahler potential. For nonsupersymmetric vacua the
results were less detailed, mainly because the constraint $V'=0$ is
quadratic in $F$, and, unlike the situation in \cite{Denef:2004cf},
no regime exists in which the equations can be linearized. To make
further progress, we will now study a class of model K\"ahler
potentials for which all vacua can be computed explicitly.

In general, at large volume, the K\"ahler potential is given by
(\ref{Kpot}): $K = -3 \log (4 \pi^{1/3} V_X)$, where $V_X$ is the
volume of $X$ regarded as a function of the moduli $s_i$. Unlike the
case of a Calabi-Yau, where the volume function is always a third
order homogeneous polynomial in the K\"ahler moduli, no strong
constraints on $V_X$ are known for $G_2$ holonomy manifolds --- just
that the volume function is homogeneous of degree $7/3$ and that
minus its logarithm is convex, i.e.\ the second derivative of $K$,
which gives the kinetic energies of the moduli, is positive
definite. In general it is difficult to find simple candidate volume
functions which satisfy this positivity constraint. The most general
homogeneous degree $7/3$ function is of the form
\begin{equation}\label{RFvolgen}
   V_X =\prod_{k=1}^n s_k^{a_k} f(s_i)
\end{equation}
with $\{a_k\}$ such that
\begin{equation}
 \sum_{k=1}^n a_k=\frac{7}{3}.
\end{equation}
and $f(s_i)$ invariant under scaling. If we now suppose that we are
in a region of moduli space where $f(s_i)$ is approximately constant
then we can take
\begin{equation}\label{RFvol}
   V_X =\prod_{k=1}^n s_k^{a_k},
\end{equation}
and this in fact gives a positive moduli space metric. This
justifies this particular choice of Kahler potentials.

The above choice of $V_X$ gives a simple geometry to the moduli
space which is quite natural. The K\"ahler metric is
\begin{equation}
   ds^2 = \sum_{i=1}^n \frac{3a_i}{4s_i^2} dz_i d\bar{z}_i
        = \sum_{i=1}^n \frac{3a_i}{4s_i^2} (dt_i^2+ds_i^2)
\end{equation}
This is locally the metric of the product of $n$ hyperbolic planes
$\mathbf{H}^2$, which is:
\begin{equation}
   ds^2 = \frac{1}{\l^2 x^2} (dx^2+dy^2)
\end{equation}
where $\l$ is connected to the curvature tensor by
\begin{equation}
   \hat{R}_{12}=-\l^2 e_1\wedge e_2
\end{equation}
So locally the moduli space is $\mathbf{H}^{2n}$. Globally it is
given by $\mathbf{H}^{2n}/\mathbb{Z}^n$, because the axions $t_i$
are periodic variables. In this class of $n$-parameter K\"ahler
potentials labeled by $a_i$, all the information about $X$ is
contained in the values of the $a_i$. Since the moduli space metric
(equivalently moduli kinetic terms) is singular if $a_i = 0$ and the
moduli have the wrong sign kinetic terms if $a_i < 0$, we take $a_i
> 0$. We will not restrict to any other particular values for the
$a_i$ if it is not necessary to do so\footnote{
To find examples which realise these Kahler potentials,
consider the case $n=7$ and $a_i = 1/3$, Then this Kahler potential correctly
describes the seven radial moduli of $X = T^7$ and certain orbifolds thereof
\cite{lukas}.}.

The potential on the moduli space is given by (\ref{Vsplit}):
\begin{equation}\label{RpotentialF}
   V = \frac{c_2^2}{48 \pi \, V_X^3} \biggl(
                3+\sum_{j=1}^n a_j \a_j s_j(\a_j s_j -3) \biggr)
        +\frac{1}{48 \pi\, V_X^3}(\vec{N}\cdot\vec{t}+c_1)^2.
\end{equation}
where $\a_j \equiv -\frac{N_j}{c_2 a_j}$.

\subsection{Description of the Vacua} \label{sec:descrvac}

We now describe the vacua ie the critical points of $V$.
The equations for the axions give
\begin{equation}\label{RSolEq2}
   \vec{N} \cdot \vec{t}+c_1=0
\end{equation}
which fixes this particular linear combination of axions. We are not
concerned with fixing the remaining axions, since they are compact
fields and will be fixed by any non-perturbative corrections. Our
interest is in the moduli $s_i$. The equations of motion for the
$s_i$ reduces to a system of $n$ quadratic equations. For the case
at hand these are equivalent to:
\begin{equation}\label{RSolEq3}
   \sum_{j=1}^n 3a_j h_j(h_j-3)-2h_i^2+3h_i+9=0,
\end{equation}
where we defined $h_j\equiv \a_j s_j$  (no sum). Note that this
system separates in $n$ quadratic equations in one variable $h_i$.
The solutions are therefore of the form
\begin{equation} \label{solhi}
 h_i = \frac{3}{4} + m_i H
\end{equation}
where $m_i=+1$ or $-1$, and $H$ is determined by substituting this
in (\ref{RSolEq3}). This results in a single quadratic equation:
\begin{equation}\label{RSolEq8}
    5 H^2 -\frac{9}{2} A H -\frac{27}{16}=0.
\end{equation}
where
\begin{equation}
 A \equiv \vec{a}\cdot\vec{m}.
\end{equation}
A priori therefore, $H$ can take two possible values:
\begin{equation}\label{RSolEq12}
   H^{\pm}_{(\vec{m})}=\frac{3}{20} \left( 3A \pm
                   \sqrt{9 A^2+15} \right)
\end{equation}
However, because
\begin{equation}
   H^+_{(\vec{m})}= - H^-_{(-\vec{m})}
\end{equation}
we only ever need to consider, say, the negative branch of the
square root to get all solutions in (\ref{solhi}). In total,
therefore, the number of vacua for a fixed choice of fluxes is
$2^{n}$.
We choose the following parametrisation: take all $2^n$ choices for
$\vec{m}$. Then
\begin{equation} \label{hsol}
 h^{(\vec m)}_i = {3 \over 4} + m_i H_{(\vec m)}
\end{equation}
with $H \equiv H^-$.

We can thus think of the vacua as the states of a system with $n$
``spins'' $m_i$. When all spins are aligned with the ``external
field'' $\vec{a}$, that is if all $m_i=+1$, we have
\begin{equation}
 A = 7/3, \quad h_i = {3 \over 5}
\end{equation}
When all spins are anti-aligned ($m_i=-1$), this becomes
\begin{equation}
 A = -7/3, \quad h_i = 3.
\end{equation}
The first of these can be shown to be the supersymmetric AdS vacuum
discussed in \cite{Acharya:2002kv} whilst the second is the unstable
de Sitter vacuum also discussed there. The remaining $2^n-2$ are all
non-supersymmetric and could be either de Sitter or anti de Sitter.
The metastability of these vacua will be analyzed in section
\ref{sec:metastab}.

\begin{figure}
\begin{center}
  \epsfig{file=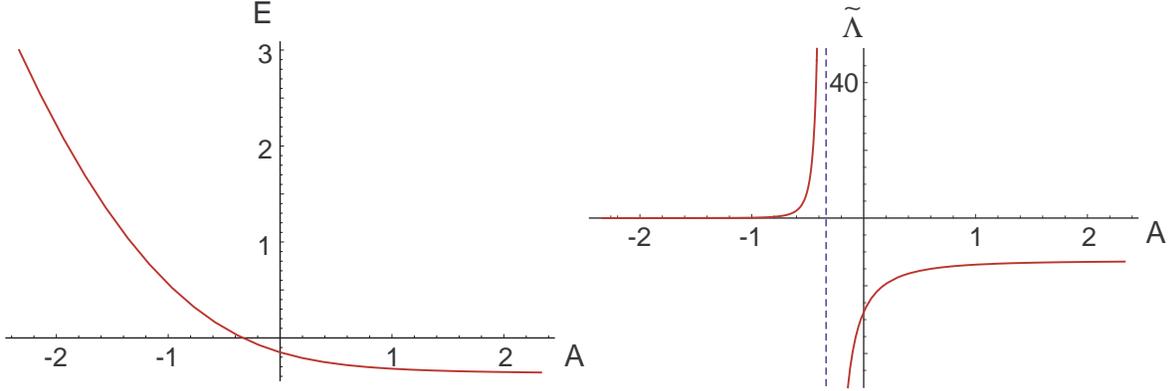,height=6cm,angle=0,trim=0 0 0 0}
  \caption{\small \emph{Left}: $E$ as a function of $A \equiv
  \vec{a}\cdot\vec{m}$. Here $E(-7/3)=3$ and $E(7/3)=-9/25$.
  \emph{Right}: dependence of $\tilde{\Lambda}$ on $A$. At $A=-7/3$, $\tilde{\Lambda} \approx 10^{-3}$,
  and at $A=7/3$, $\tilde{\Lambda} \approx -13$. The divergence at $A=-1/3$ is due to the vanishing of
  the volume there.}
  \label{vacE}
\end{center}
\end{figure}
Substituting (\ref{hsol}) in (\ref{RpotentialF}), we get that the
energy of these vacua is given by $V = \frac{c_2^2}{48 \pi V_X^3} E$
with\footnote{The normalization is chosen such that $E$ equals the
term inside the big brackets in (\ref{RpotentialF}). Also, $c_2^2 E
= \frac{3}{4} \left(|F|^2 - 3|W|^2\right)$.} $E = \frac{2}{3} H^2 -
\frac{3}{8}$. At fixed volume, the vacuum energy varies only through
$E$. The dependence of $E$ on $A$ is shown on the left in fig.\
\ref{vacE}. However, the volume depends on $\vec{m}$ as well:
\begin{equation} \label{volexpr}
 V_X = \prod_{i=1}^n s_i^{a_i} = \prod_{i=1}^n \left( \frac{c_2
 a_i}{|N_i|} \right)^{a_i} \prod_{i=1}^n |h_i|^{a_i},
\end{equation}
so the total vacuum energy is, up to $\vec{m}$-independent factors:
\begin{equation} \label{tildeLambda}
 \Lambda \sim \tilde{\Lambda} \equiv E \, \prod_{i=1}^n |h_i|^{-3a_i} =
 E \, \bigl|\frac{3}{4} + H\bigr|^{-\frac{7}{2}-\frac{3A}{2}} \,
 \bigl|\frac{3}{4} - H\bigr|^{-\frac{7}{2}+\frac{3A}{2}}.
\end{equation}
The dependence of this on $A$ is shown on the right in fig.\
\ref{vacE}. The divergence at $A=-1/3$ is due to the vanishing of
the volume there, as all $h_i$ with $m_i=+1$ vanish at this point.
Obviously the supergravity approximation breaks down in this regime.
One notable fact is further that the smallest positive cosmological
constant is obtained when all spins are down, while the smallest
negative cosmological constant (in absolute value) is obtained when
all spins are up, i.e.\ at the supersymmetric critical point.

\begin{figure}
\begin{center}
  \epsfig{file=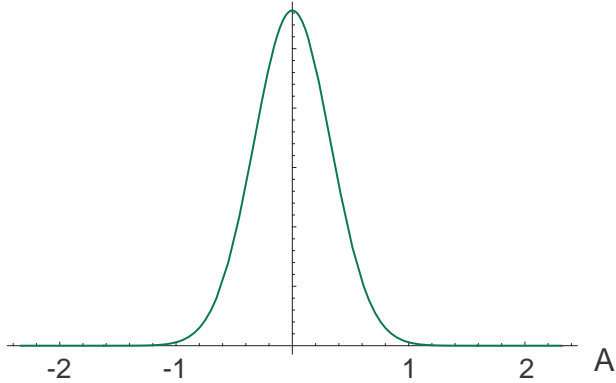,height=6cm,angle=0,trim=0 0 0 0}
  \caption{\small Distribution of $A$ values for $a_i=7n/3$, $n=50$.}
  \label{Adistrplot}
\end{center}
\end{figure}

At large $n$, the vast majority of vacua will be ``halfway'' the
extrema. More precisely, if say all $a_i = a = 7/3n$, the variable
$A$ will be binomially distributed around $A=0$, as illustrated in
\ref{Adistrplot}. At large $n$ this distribution asymptotes to the
continuous normal density
\begin{equation} \label{Adistr}
 d \CN[A] \approx \frac{2^n}{\sqrt{2\pi} \sigma} \exp\bigl(-\frac{A^2}{2 \sigma^2}\bigr) \,
 dA; \qquad \sigma=\frac{7}{3 \sqrt{n}}.
\end{equation}
For large $n$ this is sharply peaked around $A=0$, for which
$E=-3/20$ and $|F|^2 = \frac{119}{80} c_2^2$, so the majority of
vacua are AdS and break supersymmetry at a scale $M_{\rm susy}^2 =
|F|/V_X^{3/2} \sim c_2 M_p^3/m_p$.

Thus far we have considered solutions in terms of $h_i$. The actual
values of the moduli are given by
\begin{equation} \label{sfromh}
 s_i = -\frac{c_2 a_i h_i}{N_i}.
\end{equation}
Since the moduli fields $s_i$ are {\it positive} in the supergravity
approximation, the signs of the $h_i$ and flux quanta $N_i$ must be
correlated. Without loss of generality we take $c_2$ to be positive.
Then $N_i$ and $h_i$ must have opposite signs. Now as long as
$A>-1/3$, every $h_i$ is automatically positive, so any such vacuum
must have negative $N_i$. When $A<-1/3$, $h_i$ is positive if
$m_i=-1$ and negative if $m_i=+1$, so these vacua must have
$\mbox{sign } N_i = \mbox{sign } m_i$. Recall that the condition
$A<-1/3$ is also the condition to have $E>0$.

Thus, for any given $\vec{m}$, there is a unique choice of sign for
each $N_i$ which renders all $s^j$ positive for all choices of
$|N_i|$.

On the other hand, not all given, fixed fluxes $N$ gives rise to the
same number of vacua. The following cases can be distinguished:
\begin{itemize}
 \item All $N_i<0$: set of vacua = $\{ \vec{m} | A \equiv \vec{a}\cdot\vec{m}>-1/3 \} \cup \{ (-1,-1,\cdots,-1)
 \}$. All vacua in the first set are AdS. The additional one is dS
 (but is unstable, as we will discuss in the next section).
 There are of order $2^n$ such vacua (between $2^{n-1}$ and
 $2^n$ for example when all $a_i$ are equal as for the distribution
 of fig.\ \ref{vacE}).
 \item Some $N_i>0$, and $\vec{a} \cdot \mbox{sign}(\vec{N}) < -1/3$: just one
 vacuum, given by $m_i = \mbox{sign}(N_i)$. This vacuum is dS (but
 again  will turn out to be unstable).
 \item Some $N_i>0$, and $\vec{a} \cdot \mbox{sign}(\vec{N}) > -1/3$: no vacua.
\end{itemize}

As noted before, the choices of $\vec{m}$ for which $A$ is at or
near $-1/3$ do not correspond to vacua within the region of validity
of our computations for reasonable values of $c_2$,
because some of the moduli, and hence the
volume, will be at or near zero then.




\subsection{Metastability} \label{sec:metastab}

We now turn to the question of metastability. We found that all dS
vacua (i.e.\ the vacua with $A<-1/3$) for our model ensembles have a
tachyon and hence are perturbatively unstable. The proof can be
found in appendix \ref{app:metastab}. We therefore focus on AdS
vacua in what follows.

In general the condition for the perturbative stability of an AdS
vacuum is the Breitenlohner-Freedman bound
\cite{Breitenlohner:1982bm}, which is given by
\begin{equation} \label{RStabCond1}
   \hat{\partial}_{i}\hat{\partial}_{j} V-\frac{3}{2} V \delta_{ij} \geq 0
\end{equation}
where the derivatives are with respect to the canonically normalised
scalars, or equivalently the Hessian is expressed in an orthonormal
frame.

The details of the stability analysis are somewhat technical and are
given in appendix \ref{app:metastab}. We find that exponentially
large numbers of the $2^n$ vacua are in fact metastable.
Specifically, vacua for which $A > {1 \over 3}$ are always
metastable. Vacua with $-{1 \over 3} < A < {1 \over 3}$ can in
principle also be metastable (and actually turn out to be local
minima), but this is rather exceptional: they correspond to having
only one of the $m_i$ equal to $+1$. In particular, since $\sum_i
a_i = 7/3$, this means that the corresponding $a_i$ must be greater
than $1$, and thus there cannot be more than two such solutions.

Some lower bounds on the numbers of metastable vacua with $A \geq
1/3$ can be derived as follows. For simplicity, but without loss of
generality, we put
\begin{equation}\label{Rorderai}
   a_n \geq a_{n-1} \geq \ldots \geq a_1.
\end{equation}
When $a_n\geq \frac{4}{3}$, all solutions with $m_n=+1$ correspond
to $A\geq \frac{1}{3}$, and we have
\begin{equation}
    N_{\rm stab} \geq 2^{n-1}
\end{equation}
one of which is the supersymmetric solution.
When $a_n+a_{n-1}\geq \frac{4}{3}$ the number of stable vacua is at least
\begin{equation}
    N_{\rm stab} \geq 2^{n-2}
\end{equation}
and so on. So in a model with $a_n+a_{n-1}+\ldots+a_{n-j+1}\geq
\frac{4}{3}$, but with $a_n+a_{n-1}+\ldots+a_{n-j} < \frac{4}{3}$,
the number of stable vacua is at least
\begin{equation}
    N_{\rm stab}\geq 2^{n-j}
\end{equation}
Because of (\ref{Rorderai}) and the fact that $\sum a_i = {7 \over 3}$,
we cannot have $j>4n/7$. So for a model
with $n$ moduli, the number of stable vacua is surely bigger than
\begin{equation}
   N_{\rm stab} \geq 2^{n-4n/7} = 2^{3n/7},
\end{equation}
which is exponentially smaller than $2^n$ but still exponential in
$n$.

Actually this number is very hard to reach and for generic models
the number of stable vacua is much bigger than this. Take for
example the case $a_i=3/7n$ of the figure, for which
$\tilde{A}=\frac{3}{7} A$ is distributed according to (\ref{Adistr})
in the large $n$ limit. The number of vacua with $A>1/3$ is then
given by integrating the distribution (\ref{Adistr}) from $A=1/3$ to
$A=\infty$. At large $n$ this gives asymptotically
\begin{equation}
 N_{\rm stab}/2^n \approx \frac{7}{\sqrt{2 \pi n}} \exp(-n/98).
\end{equation}
Again for large $n$ this is an exponentially small fraction, but
still exponentially large in absolute number. In fact for $n=100$
the stable fraction is still about $10\%$. For $n=1000$, this goes
down to about $10^{-6}$, but this is not a small number compared to
the total number of vacua, which is $2^{1000} \sim 10^{300}$.

\subsection{Distributions over moduli space}

Let us fix $c_2$ and $\vec{m}$. We want to study the distributions
of physical quantities over the space of vacua parametrized by
$\vec{N}$. As discussed in section \ref{sec:descrvac}, the sign of
each $N_i$ is completely determined by $\vec{m}$. We can therefore
restrict to counting positive $\tilde{N}_i \equiv |N_i|$.

Let us start by finding the number of such vacua in a region
$\mathcal{R}$ given by $s_i\geq s_i^*$. By the substitution $s_i=
c_2 a_i |h_i^{(\vec m)}|/\tilde{N}_i$, this condition becomes
\begin{equation}
   \tilde{N}_i \leq \frac{c_2 a_i |h_i|}{s_i^*}.
\end{equation}
So, in the large N approximation, the number of vacua at fixed
$\vec{m}$ in this region is
\begin{equation}\label{RNsmag}
   \mathcal{N}_{(\vec m)}(s_i\geq s_i^*) = c_2^n
    \prod_{i=1}^n \frac{a_i |h_i|}{s_i^*} = \left(\frac{4c_2}{3}\right)^n
    \prod_{i=1}^n |h_i| \, \mbox{vol}(\hat{\mathcal{R}}).
\end{equation}
As in (\ref{susycount}), $\hat{\mathcal{R}}$ is the region of the
complexified moduli space projecting to $\CR$. In particular, the
number of vacua in any finite region of moduli space is finite. Note
that in the supersymmetric case $h_i=3/5$, this reproduces the
general formula (\ref{susycount}). More generally, we see that also
nonsupersymmetric vacua are distributed uniformly with respect to
the volume form in the supergravity approximation, but that their
density relative to the supersymmetric vacua, given by $\prod_i (5
h_i/3)$, is higher. Moreover, the density grows with increasing
numbers of anti-aligned spins. The highest density is that of the dS
vacua with all $m_i=-1$, which is $5^n$ higher than the density of
supersymmetric vacua. Obviously, this does \emph{not} mean that in
total there are $5^n$ times more dS maxima as supersymmetric
critical points, since we know there is a one-to-one correspondence
between them (in the supergravity approximation). The density in a
given region is higher simply because the nonsupersymmetric vacua
sit at larger radii. Integrated over the entire moduli space in the
supergravity approximation, we do not run into a paradox, because
both numbers are then infinite. In the fully quantum corrected
problem, the numbers presumably will be finite (by analogy of what
happens for type II flux vacua due to worldsheet instanton
corrections), but then of course also the relative densities will
change.

\vspace{5mm}

In order to have a meaningful four dimensional effective theory,
decoupled from the KK modes, we need the Kaluza-Klein radius to be
smaller than the AdS radius. Taking all $s_i \sim s$, we have
\begin{eqnarray}
 R^2_{\rm AdS} &\sim& \frac{m_p^2}{\Lambda} \sim \frac{V_X^3}{c_2^2 m_p^2} \sim \frac{s^7}{c_2^2 m_p^2} \\
 R^2_{\rm KK} &\sim& \frac{s^{2/3}}{M_p^2} \sim \frac{s^3}{m_p^2},
\end{eqnarray}
so $R_{\rm KK} < R_{\rm AdS}$ iff $s > c_2^{1/2}$.\footnote{The
assumption that all $s_i \sim s$ can be relaxed. Then one can prove
that $R_{\rm KK} < R_{\rm AdS}$ is guaranteed if $s_i>c_2^{4/7}$.
However for most vacua, $s_i>c_2^{1/2}$ will be sufficient to have
the required scale hierarchy, so we stick to this estimate.}

This also ensures the validity of the supergravity approximation.
The number of vacua at fixed $\vec{m}$ satisfying this condition is
given by (\ref{RNsmag}):
\begin{equation}\label{RNsmag2}
   \mathcal{N}_{(\vec m)}(s_i\geq c_2^{1/2}) = c_2^{n/2}\prod_{i=1}^n a_i
   |h_i|
\end{equation}

\vspace{5mm}

Finally, to get the total density for all possible $\vec{m}$ as
well, we must sum (\ref{RNsmag}) over all $\vec{m}$. Because of the
absolute values and the nonlinear dependence of $h_i$ on $\vec{m}$,
this is not easy to compute analytically even in special cases. But
one can get numerical results without much effort. For example in
the case with all $a_i=7/3n$, we get numerically that
\begin{equation}
 \sum_{\vec m} \prod_i |h_i| \approx (3.328)^n.
\end{equation}
This approximation becomes better for large $n$, but is quite good
for smaller values as well. For $n=1$, the exact result is
$3+3/5=3.6$, which is already not too far from this expression.

\subsection{Distributions of volumes and cosmological constants}

\subsubsection{Fixed $\vec{m}$}

Now, let us consider the distributions of the volumes $V_X$. 
For a given $\vec{m}$ the volume is given by (\ref{volexpr}):
\begin{eqnarray}
   V_X &=& {c_2}^{7/3}
    \prod_{j=i}^n\biggl(\frac{a_j |h_j| }{\tilde{N}_j}\biggr)^{a_j}
\end{eqnarray}
We see that as the $\tilde{N}_j$ go to infinity, $V_X$ tends to
zero, so the density of vacua (strongly) increases with decreasing
volume, i.e.\ large volumes are suppressed. This agrees with what we
found earlier for the general supersymmetric case in section
\ref{sec:largevolsuppr}.

\begin{figure}
\begin{center}
  \epsfig{file=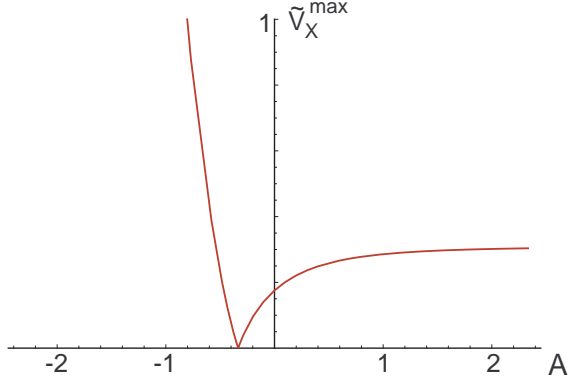,height=6cm,angle=0,trim=0 0 0 0}
  \caption{\small Dependence of $\tilde{V}_X^{\rm max}$ on $A$. At $A=7/3$,
  $\tilde{V}_X^{\rm max} = (3/5)^{7/3} \approx 0.3$, at $A=-7/3$, $\tilde{V}_X^{\rm max} = 3^{7/3} \approx 13$, and
  the zero is at $A=-1/3$.}
  \label{vols}
\end{center}
\end{figure}

The maximal value that $V_X$ can assume is obtained when all
$\tilde{N}_j=1$:
\begin{equation} \label{modvolbound}
  V_X^{\rm max} = c_2^{7/3}\prod_j a_j^{a_j} |h_j|^{a_j}
  \equiv c_2^{7/3} \prod_j a_j^{a_j} \, \tilde{V}_X^{\rm max}.
\end{equation}
Here we isolated the $\vec{m}$-dependent product $\prod_j
|h_j|^{a_j}$:
\begin{equation} \label{tVmaxdef}
 \tilde{V}_X^{\rm max} = \bigl|\frac{3}{4} + H\bigr|^{\frac{7}{6}+\frac{A}{2}} \,
 \bigl|\frac{3}{4} - H\bigr|^{\frac{7}{6}-\frac{A}{2}}.
\end{equation}
The variation of $V_X^{\rm max}$ over different choices of $\vec{m}$
is entirely given by the dependence of this function on
$A=\vec{a}\cdot\vec{m}$. This is shown in fig.\ \ref{vols}.

When we take all $a_i = 7/3n$, and we consider say the susy case
$m_i=+1$ so $h_i = 3/5$, (\ref{modvolbound}) becomes
\begin{equation}
 V_X^{\rm max}|_{\rm susy} = \biggl( \frac{7 \, c_2}{5 \, n} \biggr)^{7/3},
\end{equation}
which is similar to the toy model result (\ref{volbound}).



Using the usual continuum approximation for the fluxes, it is
possible to get explicit expressions for the volume distribution of
vacua at fixed $\vec{m}$. Taking as example again the case
$a_i=7/3n$, we show in appendix \ref{gamma} that for $v \equiv
V_X/V_X^{\rm max} \leq 1$, at fixed $\vec{m}$ (or fixed $A$), the
vacuum number density is
\begin{equation} \label{vdistr}
 d \CN_A[v] = \frac{\left(\frac{3n}{7}\right)^n}{(n-1)!} (-\ln
 v)^{n-1} \, v^{-\frac{3n}{7}-1} \, dv.
\end{equation}
Therefore the total number of such vacua with $V_X \geq V_*$ goes as
$V_*^{-3n/7}$, in agreement with the general estimates of section
\ref{sec:largevolsuppr}. Note that in addition here, the density of
vacua vanishes to order $(n-1)$ near the cutoff $v=1$.


Thus, in accordance with general expectations, we see that large
volumes are strongly suppressed, and more so when there are more
moduli.

Analogous considerations can be made about the distribution of
cosmological constants at fixed $\vec{m}$:
\begin{equation}
   \Lambda = \frac{c_2^2 \, E}{48 \pi V_X^3},
\end{equation}
with $E(A)$ as defined in section \ref{sec:descrvac}. Clearly, the
distribution of $\Lambda$ is completely determined by the
distribution of $V_X$. Thus, because large volumes are suppressed,
we see that small cosmological constants are suppressed. In
particular there is a lower bound on $|\Lambda|$:
\begin{equation}
 |\Lambda|_{\rm min} = \frac{1}{48 \pi} \, c_2^{-5} \prod_j a_j^{-3a_j} \, \frac{E}{(\tilde{V}_X^{\rm
 max})^3} = \frac{1}{48 \pi} \, c_2^{-5} \prod_j a_j^{-3a_j} \,
 \tilde{\Lambda}
\end{equation}
with $\tilde{\Lambda}$ as given by (\ref{tildeLambda}) and plotted
in figure \ref{vacE}.

\subsubsection{All $\vec{m}$}

So far in this section, we have studied distributions at fixed
$\vec{m}$, or equivalently at fixed $A$. To get complete statistics
of all vacua, we need to combine these results with the distribution
of solutions over values of $A$.

One general observation one can make is that since there are no
metastable vacua at $A \leq -1/3$, the largest possible volume of a
metastable vacuum is obtained at $A=7/3$ (and $N_i=1$), the
supersymmetric solution. This can be seen from fig.\ \ref{vols}.
Therefore, the maximal volume for a metastable vacuum is $c_2^{7/3}
\, \prod_j a_j^{a_j} \, (3/5)^{7/3}$. It is not hard to show further
that $(7/3n)^{7/3} \leq \prod_j a_j^{a_j} \leq (7/3)^{7/3}$. The
former corresponds to all $a_i$ equal, the latter to the limiting
case $a_i \to 0$ for all but one $a_i$. Similar considerations hold
for the cosmological constant. Thus we arrive at the result that for
our ensembles:
\begin{equation}
 V_X \leq 2.2 \, c_2^{7/3} \, \hat{n}^{-7/3}, \qquad
 |\Lambda| \geq 2.3 \times 10^{-4} \, c_2^{-5} \, \hat{n}^7
\end{equation}
where $1 \leq \hat{n} \leq n$.

To get more refined results on the actual distributions, we need the
precise distribution of solutions over $A$. This depends on the
values of $a_i$. When all $a_i=7/3n$ for example, the distribution
is binomial, peaked around $A=0$, which for sufficiently large $n$
is well approximated by the normal distribution given in
(\ref{Adistr}). When all $a_i$ are approximately equal, the
distribution will still be approximately binomial, and
(\ref{Adistr}) is still a good approximation for large $n$. When say
$a_1=1$ and all other $a_i=4/3n$ with $n$ large, the distribution
will have two peaks, one at $A=1$, corresponding to $m_1=+1$, and
the other one at $A=-1$, corresponding to $m_1=-1$. In the following
we will work with the distribution given in (\ref{Adistr}).

\begin{figure}
\begin{center}
  \epsfig{file=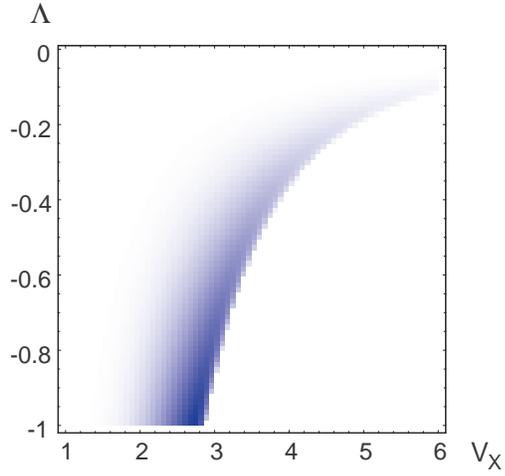,height=7cm,angle=0,trim=0 0 0 0}
  \caption{\small Density plot of the joint distribution of AdS vacua over $V_X$ and $\Lambda$, for
  $c_2=100$, $n=20$.}
  \label{ccvol}
\end{center}
\end{figure}

The joint distribution for $A$ and $V_X$ is then given by
multiplying the distributions (\ref{Adistr}) and (\ref{vdistr}).
Note that (\ref{vdistr}) depends on $A$ through $V_X^{\rm max}$. One
can also change variables from $A$ to $E$ and thus write down a
joint distribution for $E$ and $V_X$, or equivalently (and
physically more relevantly) for $\Lambda$ and $V_X$, as we did for
the Freund-Rubin ensemble. The explicit expressions are not very
illuminating, so we will not get into details here. An example is
plotted in fig.\ \ref{ccvol}.

Clearly, $V_X$ and $\Lambda$ are correlated, as in the Freund-Rubin
case. This follows directly of course from the relation $\Lambda
\sim E/V_X^3$. In particular given one variable, we get a roughly
Gaussian distribution of the other variable. Qualitatively this is
somewhat similar to our Freund-Rubin ensemble
(compare\footnote{Notice that fig.\ \ref{vacpoints} shows
$|\Lambda|^{-1}$ instead of $\Lambda$ on the vertical axis.} with
fig.\ \ref{vacpoints}), although the details are different. For
example in the Freund-Rubin case, at fixed $\Lambda$, the vacua
accumulate near the lower bound on $V_X$, whereas in the $G_2$ case
they accumulate near the upper bound. Without constraint on
$\Lambda$ this gets reversed for both.

\begin{figure}
\begin{center}
  \epsfig{file=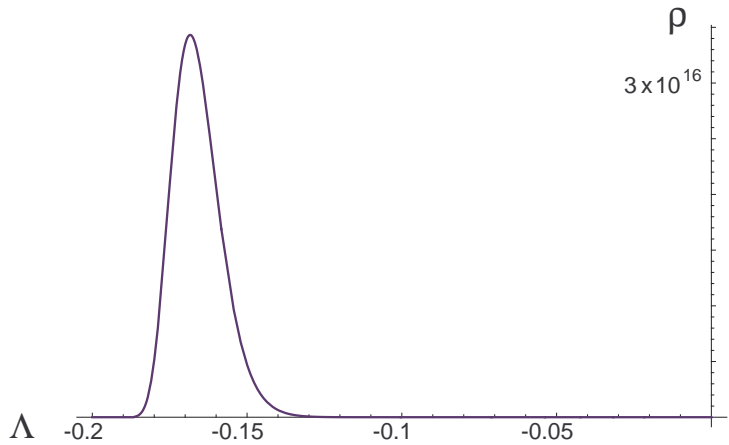,height=7cm,angle=0,trim=0 0 0 0}
  \caption{\small Distribution of cosmological constants $d\CN = \rho(\Lambda) \, d\Lambda$.
  Here we took $a_i=7/3n$, $n=50$, $c_2=100$, and we restricted to stable vacua
  with $V_X \geq 5$.}
  \label{ccdistr}
\end{center}
\end{figure}

Finally, one can get the distribution of cosmological constants for
vacua with volume above some cutoff value by integrating the joint
density over $V_X$. Again for the case with all $a_i=7/3n$, we
obtain the distribution for $\Lambda$ shown in fig.\ \ref{ccdistr}
for $n=50$, $c_2=100$ and $V_X \geq 5$. The cutoff at smaller values
of $|\Lambda|$ appears because of the lower cutoff we impose on the
volume; the lower we take the volume cutoff, the lower the value of
$\Lambda$ at which the density peaks. This is because $1/V_X^3$ sets
the scale for $\Lambda$. The cutoff for small $|\Lambda|$ appears
because the ensmemble does not contain vacua with arbitrarily large
volume, as discussed at length before.

\subsection{Supersymmetry breaking scales}

Let us consider the distribution of the supersymmetry breaking scale, which for a fixed $\vec{m}$
is given by
\begin{eqnarray}
    M_{{\rm susy}(\vec{m})}^2 &=& \left(\sqrt{e^K g^{i\bar{j}}DW_i\overline{DW_j}}\right)_{\vec{m}}\nonumber\\
             &=& \frac{c_2}{(48\pi)^{1/2}}\frac{G_{\vec{m}}^{1/2}}{V_X^{3/2}}
\end{eqnarray}
where
\begin{equation}
    G_{\vec{m}}\equiv \left(\frac{7}{3}+\frac{9}{4}A^2
    \right)H^2+\frac{15}{8} A H+\frac{21}{64}
\end{equation}
$G_{\vec{m}}$ is a decreasing function of $A$, which
is positive for $-\frac{7}{3}\leq A < \frac{7}{3}$ and zero in the supersymmetric solution
($A=\frac{7}{3}$).

We see that the expression for $M_{\rm susy}^4$ is the same as that
of the cosmological constant, but with $G_{\vec{m}}$ instead of
$E_{\vec{m}}$.  So the distribution of the supersymmetry breaking
scale at fixed $\vec{m}$ is very similar to that of the cosmological
constant:
\begin{itemize}
    \item it is completely determined by the volume distribution;
    \item low supersymmetry breaking scales are suppressed (in line with the general expectations of section
    \ref{sec:gensusybreakingscales} \footnote{although the situation here is a bit different
    then the case on which we focused there, namely $\Lambda \sim 0$, which cannot be obtained in the
    present ensemble.});
    \item $M_{\rm susy}^2$ has a lower bound given by
        \begin{equation}
           (M_{\rm susy}^2)_{\rm min}=\frac{1}{(48\pi)^{1/2}} \, c_2^{-5/2} \prod_j a_j^{-3a_j/2} \,
           \frac{G^{1/2}}{(\tilde{V}_X^{\rm max})^{3/2}}
        \end{equation}
  with $\tilde{V}_X^{\rm max}$ defined in (\ref{tVmaxdef}).
\end{itemize}

\begin{figure}
\begin{center}
  \epsfig{file=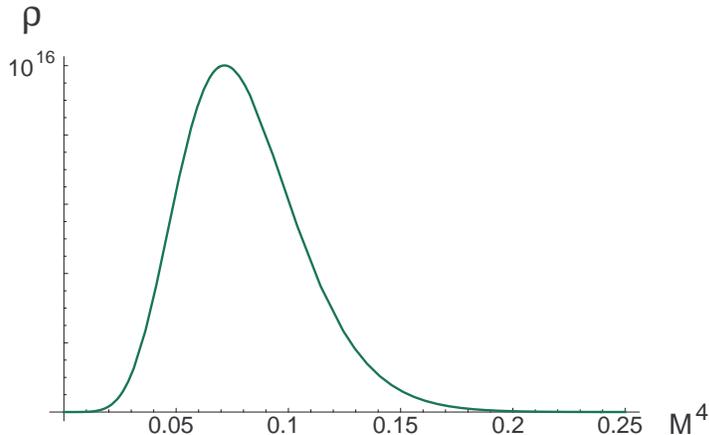,height=7cm,angle=0,trim=0 0 0 0}
  \caption{\small Distribution of supersymmetry breaking scales $d\CN = \rho(M^4) \, dM^4$, with
  $M \equiv M_{susy}/m_p$. Here we took $a_i=7/3n$, $n=50$, $c_2=100$, and we restricted to stable vacua
  with $V_X \geq 5$.}
  \label{susybrdistr}
\end{center}
\end{figure}

As for the $V_X$ and $\Lambda$ distributions, in order to get the
complete statistic of all vacua, we need to consider the
distribution of solutions over $A$, which depends on the $a_i$'s.
Similar to what we did for $\Lambda$ in the previous subsection, we
can compute the distribution of $\Lambda$ for vacua with volume
bounded by some lower cutoff. This is illustrated in fig.\
\ref{susybrdistr}. The cutoff at large values of $M_{\rm susy}$
appears because of the lower cutoff we impose on the volume, again
because $1/V_X^3$ sets the scale for $M_{\rm susy}^4$. The lower
cutoff is there because large volumes are absent.

\section{Conclusions and Discussion}


The potential importance of the emerging ideas surrounding the
landscape, e.g.\ for the notion of naturalness, is clear. These
ideas should therefore be tested {\it both} experimentally by
verifying specific predictions {\it and} theoretically within the
framework of string theory. To make progress in the latter, {\it one
should further scrutinize} the proposed ensembles of string theory
vacua, to establish whether or not these vacua truly persist after
taking into account the full set of subtle consistency requirements,
quantum corrections, and cosmological constraints
\cite{Banks:2004xh}. Parallel to that one should develop techniques
to analyze large classes of (potential) vacua without actually
having to go through their detailed constructions. Only with both
efforts combined can one hope to get to a satisfactory analysis of
the implications of the landscape picture within string theory. If
nothing else, the latter program will teach us a lot about what kind
of physics is possible in string theory and assist in guiding
explicit model builders, as our example in the introduction
demonstrates.

The focus of our paper has been on the second program. To this end,
we analyzed the statistics of Freund-Rubin and $G_2$ flux
compactifications of $M$ theory, and compared this to known IIB
results.

The statistics of Freund-Rubin vacua is very different from that of
flux compactifications on special holonomy manifolds. Most notably,
they can have arbitrarily high compactification volume, and
accumulate near zero cosmological constant. On the other hand, these
vacua {\it typically} do not have a large gap between the KK scale
and the four dimensional AdS curvature scale, so they are not really
compactifications in the usual sense. This is no longer true
\cite{Acharya:2003ii} if (in units in which the scalar curvature is
one) the Einstein metric has all length scales much smaller than 1,
but it remains a challenge to construct such manifolds with more
than a modest scale hierarchy. Alternatively, one can imagine adding
positive energy sources to lift the cosmological constant, perhaps
even to small positive values, but no viable controlled scenario
that would accomplish this is presently known. The problem is clear:
in order to lift the cosmological constant to a positive value, one
would have to add an energy source at or near the Kaluza-Klein
scale, which makes 4d effective field theory unreliable and could
easily destabilize the compactification. Adding such effects could
also drastically change the distributions of vacua over parameter
space. Therefore, the results we obtained for Freund-Rubin vacua
should not be interpreted as showing there are parts of the
landscape \emph{compatible with rough observational requirements}
that for example strongly favor large volumes. But our results do
show that this is possible in principle, and constructing
Freund-Rubin-like vacua which overcome the above mentioned problems
would therefore be all the more interesting.

The statistics of $G_2$ flux vacua has a number of universal
features. One is that the distribution of vacua over moduli space is
uniform with respect to the K\"ahler metric on moduli space. In
essence,  because the large volume region of moduli space is small
when its dimension $b_3$ is large, this implies that large
compactification volumes are suppressed, and strongly so if $b_3$ is
large. In particular there is an upper bound to the volume in a
given ensemble, set by the Chern-Simons invariant $c_2$. This is
true as well for IIB flux ensembles studied thus far and for their
presumed IIA mirror counterparts, where the maximal size is set by
the D3 tadpole cutoff $L$. Since both the IIB and the $G_2$
ensembles contain flux degrees of freedom that are not dual to flux
degrees of freedom in the other ensemble, but are possibly dual to
discrete geometrical deformations away from special holonomy, this
suggests this behavior will persist when extending the ensembles to
also sample non-flux discrete compactification data away from
special holonomy. Our results on Freund-Rubin vacua on the other
hand show that departure from special holonomy can produce the
opposite behavior. So before general conclusions can be drawn, this
question needs to be investigated in more general ensembles. Since
the full 10d geometries of such extended ensembles tend to be hard
to describe, an effective four dimensional field theory approach, as
worked out in \cite{Grimm} and
used for example in \cite{Kachru:2004jr} to find $\CN=2$ IIA flux
vacua, would probably be the best way to make progress in this
direction.

Essentially because of the limited discrete tunability of $G_2$ flux
vacua, the scale of the cosmological constant is set by the volume,
$\Lambda \sim m_p^4/V^3$. Since large volumes are strongly
suppressed at large $b_3$, it follows that small cosmological
constants are strongly suppressed as well. This contrasts with type
IIB ensembles, where the distribution of cosmological constants is
uniform near zero. Similarly, the scale of supersymmetry breaking is
set by $m_p^2/V^{3/2}$, so small supersymmetry breaking scales are
suppressed. For $b_3$ not too small, this remains true even when
taking into account tuning of Higgs mass and cosmological constants.
F-breaking IIB flux vacua similarly favor high susy breaking scales,
although $M^2_{\rm susy}$ can be tuned to be small there, and the
suppression was found to be independent of the number of moduli in
the regime where $M_{\rm susy}^2$ is much smaller than the
fundamental scale \cite{Denef:2004cf}. It is plausible that
extending the $G_2$ ensembles as discussed above would allow the
supersymmetry breaking scale to be tuned small as well, reproducing
the statistics of the generic ensembles of \cite{Denef:2004cf} in
this regime, but this would still favor higher scales. Another
generalization which we did not consider in this work is D-term
supersymmetry breaking. We refer to \cite{Denef:2004cf} for further
discussion of this possibility.


The next step to take would be to embed standard-like models in
these (and other) flux ensembles, and study the distributions of
various couplings. In particular it would be interesting to analyze
the impact of environmental constraints on these distributions, to
see if for example the split supersymmetry scenario of
\cite{Arkani-Hamed:2004fb} naturally emerges, and if these ensembles
are ``friendly neighborhoods'' of the landscape, in the terminology
of \cite{Arkani-Hamed:2005yv}. To get an idea of the possibly very
important role of environmental constraints in determining the
physically relevant distributions, consider the ``atomic principle''
\cite{Arkani-Hamed:2004fb}, which, when the light Yukawa couplings
are fixed, constrains the weak scale $v$ to lie in a narrow window
set by the (naturally small) QCD scale. More generally, leaving the
Yukawa couplings $y$ arbitrary, it is actually the combination $y v$
which is constrained to lie in a narrow window, so the environmental
explanation of the smallness of $v$ would be lost if $y$ scans
densely around zero in the ensemble. Now, in $M$ theory vacua where
the quarks and leptons are localized at isolated points in the extra
dimensions \cite{AW,Acharya:2001gy}, the Yukawa couplings are
generated by $M$2-brane instantons wrapping 3-cycles containing
these points \cite{Acharya:2001gy} and are therefore roughly of
order $e^{-s}$, with $s$ one of the geometric moduli we have worked
with throughout our analysis. The values of $s$ vary over the
ensemble, so the Yukawas scan many orders of magnitude \cite{yuk},
but not uniformly around zero: as we have seen, large values of $s$
are suppressed and bounded above. Hence these ensembles are
precisely of the kind envisioned in \cite{Arkani-Hamed:2005yv} that
would be sufficiently friendly to allow the atomic principle to play
a role in explaining the smallness of $v$. In fact, it may even play
a role in explaining the smallness of the light Yukawas. A priori,
these ensembles favor smaller $s$ and therefore larger Yukawas.
However, with smaller Yukawas, the Higgs is allowed to vary over a
wider range. Taking this into account gives an \emph{enhancement} of
order $e^s$ in the number of vacua compatible with the atomic
principle, and this could easily overpower the a priori $1/s^k$
suppression of large $s$. Apart from illustrating the potential
power of environmental principles, this also shows that such
constraints can strongly influence the physically relevant
distributions. We plan to return to these and other issues in future
work.

\vskip2cm

{\large \sf Acknowledgements}

\vskip3mm

We would like to thank Nima Arkani-Hamed, Michael Dine, Mike
Douglas, Shamit Kachru, Greg Moore, Sav Sethi and Eva Silverstein
for valuable discussions.

\appendix

\section{Normalizations}

We defined
\begin{equation}
 K = - 3 \ln \left( 4 \pi^{1/3} V_X \right); \qquad W_{flux}(z) = {1 \over \kappa_4^3} N_i
 z^i
\end{equation}
The expression for the supergravity potential is the standard one
for the case of dimensionless\footnote{The scalars can of course be
given standard dimensions by rescaling $\phi = z / \kappa_4$, which
absorbs the $\kappa_4^2$ factor for the first term in
(\ref{sugrapotential}).} scalars:
\begin{equation} \label{sugrapotential}
 V = \kappa_4^2 \, e^K \, (g^{i\bj} D_i W \bar{D}_{\bj} \bar{W} - 3 |W|^2)
\end{equation}
Note that when expressed in terms of the moduli $z$, none of these
effective four dimensional field theory quantities explicitly
contains the fundamental scale $\ell_M$, only the four dimensional
Planck scale $\kappa_4$, which is the directly measurable scale in
four dimensions. To check our normalizations, it is sufficient to
verify the tension of a domain wall corresponding to an M5 brane
wrapped around a supersymmetric 3-cycle $\Sigma$, which is
Poincar\'e dual to the jump in flux $\Delta G$ across the wall.
Let's take $C=0$ for simplicity. In the supergravity theory, we have
\begin{eqnarray*}
 T_{DW} &=& 2 e^{K/2} |\Delta W| \\
 &=& {2 \over V_X^{3/2} 8 \sqrt{\pi} \kappa_4^3} |\Delta N_i z^i| \\
 &=& {2 \pi \over (4 \pi \kappa_4^2 V_X)^{3/2} \ell_M^3} |\int_{\Sigma} \varphi| \\
 &=& {2 \pi \over \ell_M^6} \mbox{vol}(\Sigma) \\
 &=& T_5 \mbox{vol}(\Sigma),
\end{eqnarray*}
which is indeed the correct expression in $M$ theory.

\section{Stability analysis} \label{app:metastab}

The sign of the potential for a particular solution depends on the
sign of $E$, with $E$ as defined in section \ref{sec:descrvac}. It
is negative when $A \equiv \vec{a}\cdot\vec{m}>-1/3$ and it is
positive when $A<-1/3$. We discuss these cases separately.

\subsection{AdS vacua}

An AdS critical point $s^0$ does not have to be a local minimum to
be perturbatively stable. It suffices that the eigenvalues of the
Hessian of $V$ are not too negative compared to the cosmological
constant, and more precisely that Breitenlohner-Freedman bound is
satisfied \cite{Breitenlohner:1982bm}:
\begin{equation} \label{RStabCond1app}
   \hat{\partial}_{i}\hat{\partial}_{j} V(s^0)-\frac{3}{2} V(s^0) \delta_{ij} \geq
   0,
\end{equation}
i.e.\ this matrix should be positive definite.

The derivatives are done with respect to the canonically normalized
scalars. The relevant kinetic term expanded around the critical
point $s^0$ is (in 4d Planck units):
\begin{eqnarray}
   g_{i\bar{j}}\partial_\mu z^i\partial^\mu \bar{z}^{\bar{j}} &=&
   \sum_{i}\frac{3a_i}{4(s_i^0)^2}(\partial_\mu t_i\partial^\mu t_i
   +\partial_\mu s_i\partial^\mu s_i) + \cdots
   \\
   &&\equiv \sum_{i} \frac{1}{2}(\partial_\mu \hat{t}_i\partial^\mu \hat{t}_i
   +\partial_\mu \hat{s}_i\partial^\mu \hat{s}_i) + \cdots
\end{eqnarray}
Hence the canonically normalized scalars are
\begin{equation}
 \hat{s}_i= \sqrt{\frac{3a_i}{2}}\,\frac{s_i}{s_i^0}.
\end{equation}
With this redefinition, the condition (\ref{RStabCond1app}) becomes
\begin{equation}\label{RStabCond2}
      \frac{2}{3} \frac{s_i^0 s_j^0}{\sqrt{a_i a_j}}
      \partial_i\partial_j V(s^0)-\frac{3}{2}V(s^0) \delta_{ij} \geq 0
\end{equation}
where
\begin{eqnarray}
   \partial_i\partial_j V&=&\frac{c_2^2 }{48\pi V_X^3}\left(\frac{3a_ia_j}{s_i s_j}
                           \left(3 E -2(\nu_i^2 s_i^2+\nu_j^2 s_j^2)
               +3(\nu_i s_i+\nu_js_j)\right)\right.\nonumber\\
               &&+\left.\frac{a_i}{s_i^2}\left(3 E
                 + 2\nu_i^2 s_i^2\right)\delta_{ij}\right)\\
   V           &=&\frac{c_2^2 }{48\pi V_X^3} E\nonumber\\
   E       &=& 3+\sum_{j=1}^n a_j \nu_j s_j(\nu_j s_j -3) \nonumber
\end{eqnarray}
By substituting these expressions in (\ref{RStabCond2}), using the
variables $h_i=\nu_i s_i$ and factoring out the common positive term
$\frac{c_2^2}{48\pi V_X^3}$, the stability condition becomes:
\begin{equation}\label{RStabCond3}
   M_{ij}\equiv \sqrt{a_i a_j}(6 E -4(h_i^2 +h_j^2)
      +6(h_i +h_j)) + \delta_{ij} (\frac{E}{2}
    +\frac{4}{3}h_i^2)\geq 0
\end{equation}
where
\begin{equation}
       E=\frac{2}{3}H^2-\frac{3}{8} = -\frac{3}{100}\left(5-9 A^2
                     + 3 A \sqrt{9A^2 +15}\right).
\end{equation}
Let us evaluate (\ref{RStabCond3}) on our solutions.
\begin{itemize}
  \item supersymmetric solution:
        \begin{equation}
       \mathbf{M}=\frac{54}{25} \, \mathbf{q}+\frac{3}{10} \, \mathbf{1}
    \end{equation}
    where $q_{ij}=\sqrt{a_ia_j}$ is a rank$=1$ positive definite matrix.
    We see that $\mathbf{M}\geq0$ and so that the supersymmetric solution is
    always stable.
  \item Other AdS ($A>-1/3$) solutions:
        \begin{equation}
       \mathbf{M}=Q \, \mathbf{q}+S \, \mathbf{1}+T \, \mathbf{t}
    \end{equation}
    where
    \begin{equation}
        Q = -6 E, \qquad
        S = \frac{E}{2}+\frac{4}{3} \left(H+\frac{3}{4}\right)^2, \qquad
        T = -4H
    \end{equation}
    \begin{equation}
        t_{ij}=\frac{(1-m_i)}{2} \delta_{ij}.
    \end{equation}
    The following relations are valid: $Q>0$, $T>0$ and $S+T>0$;
    so $Q\,\mathbf{q}$ and $T\,\mathbf{t}$ are always positive definite.
    $S$ has no definite sign. We have two cases:
    \begin{enumerate}
      \item when $A\geq \frac{1}{3}$, $S\geq 0$; then
        $S \, \mathbf{1}$ and consequently $\mathbf{M}$ is positive,
        and the corresponding solutions are perturbatively stable;
      \item when $-\frac{1}{3}< A  <\frac{1}{3}$, $S<0$; in this case the
            $\mathbf{M}$ matrix is not positive definite when
        $\vec{m}$ has more than one entry equal to $+1$ and the
        corresponding solutions are not stable.
        Actually, if $m_h=+1$ and $m_k=+1$, $\mathbf{M}$ has the $2\times 2$ minor
        \begin{equation}
           Q\left(\begin{array}{cc}
            a_h &\sqrt{a_ha_k}\\ \sqrt{a_ha_k}& a_k
                  \end{array}\right) +
           \left(\begin{array}{cc}
            S & 0\\ 0 & S
                  \end{array}\right)
        \end{equation}
        which has $S<0$ as one of the two egenvalues.\\
        On the other hand, when $\vec{m}$ has exactly one entry equal to $+1$, the solution
        is a local minimum and so it is stable.
    \end{enumerate}
\end{itemize}

Let us prove that when $A> -\frac{1}{3}$ and exactly one $m_i$ is
equal to $+1$, the corresponding solution is a local minimum.

    Without loss of generality we choose $i=1$.
    $\partial^2 V$ is proportional to
    $\mathbf{\hat{M}} \equiv \mathbf{q}+\frac{S'}{Q}\mathbf{1}+\frac{T}{Q}\mathbf{t}$, where
    $S' \equiv S+\frac{3}{2}E$ is negative for all considered values of $A$.
    Let us call $B=\frac{S'+T}{Q}>0$, $C=-\frac{S'}{Q}>0$, $v_i=\sqrt{a_i}$.
    So $|v|^2=7/3$, $v_1>1$ and
    \begin{equation}
       \mathbf{\hat{M}}= v\,v^t + \left(\begin{array}{cccc}
                    -C & & & \\ &B& &\\ & &\ldots&\\ & & & B
                      \end{array}\right)
    \end{equation}
    By direct study of its eigenvalues, it can be shown that this matrix is positive definite for
    $n=2$. Now we prove by induction that this matrix is positive definite for each
    $n$.

    Let us assume that $\forall y\in \IR^n$
    \begin{equation}
       y^t \mathbf{\hat{M}}_n y = (v_1 y_1 + \tilde{v}\cdot\tilde{y})^2+ B\tilde{y}^2-C y_1^2>0
    \end{equation}
    where $V=(V_1,\tilde{V})$.
    Let $u$ be the extension of $v \in \IR^n$ to $\IR^{n+1}$, i.e.\ $u_i=\sqrt{a_i}$. Then
    $\forall x\in\IR^{n+1}$ there exists $y\in\IR^n$, such that
    \begin{eqnarray}
       y_1 =x_1 & \tilde{y}\cdot\tilde{v} = \tilde{x}\cdot\tilde{u}&\tilde{y}^2= \tilde{x}^2
    \end{eqnarray}
    It follows that
    \begin{equation}
       x^t \mathbf{\hat{M}}_{n+1} x = (u_1 x_1 + \tilde{u}\cdot\tilde{x})^2+ B\tilde{x}^2-C x_1^2>0
    \end{equation}
    {\it QED}.

\subsection{dS vacua}

The solutions with positive potential (i.e.\ $A<-1/3$) must be local
minima to be metastable, and so the matrix $\mathbf{M'} \equiv Q \,
\mathbf{q}+S' \, \mathbf{1}+T \, \mathbf{t}$ (in the notations
introduced in the first part of this appendix) must be positive
definite. We now show that this is never the case.

We have $\mathbf{q}=v v^t$, where $v_i=\sqrt{a_i}$ and so
$|v|^2=7/3$. Let us consider the $n\times n$ matrix defined by
$\mathbf{M''} \equiv Q \, \mathbf{q}+S'\, \mathbf{1} + T \,
\mathbf{1}$. If $\mathbf{M'}$ is positive definite then
$\mathbf{M''}$ is positive too. Clearly, its eigenvalues are $S'+T$,
which is positive for $A<-1/3$, and
\begin{equation}
   \lambda=Q |v|^2+S'+T=\frac{7}{3}Q+S'+T.
\end{equation}
By studying $\lambda$ as a function of $A$, one finds that it is
positive for $A >-17/21$. So $\mathbf{M''}$ is not positive definite
for $A<-17/21 \approx -0.8095$ and so $\mathbf{M'}$ is not positive
for the same values.

Moreover one can study the $k\times k$ submatrix of $\mathbf{M'}$
obtained by restriction to the subspace on which $m_j=+1$ ($k$ is
the number of such $m_j$'s). This is equal to $Q \, v v^t+S'\,
\mathbf{1}_{k}$, where we are considering only $v_j$ corresponding
to $m_j=+1$. Its eigenvalues are $S'$, which is positive for
$A<-1/3$, and
\begin{equation}
   \mu=Q \sum_{i=1}^n \frac{(1+m_i)}{2}a_i + S'=
   Q(\frac{7}{6}+\frac{A}{2})+S'.
\end{equation}
By studing $\mu$ as a function of $A$, one finds that it is negative
for $\alpha<A<-\frac{1}{3}$, where $\alpha \approx -1.44075$. For
these values of $A$ we have a nonpositive minor and so $\mathbf{M'}$
is not positive definite.

By combining these two results, we see that for no value of
$A<-\frac{1}{3}$ the corresponding solution is a local minimum.
Thus, all dS critical points are unstable in our model ensembles.

\section{Volume distribution} \label{gamma}

Here we give details on how to arrive at the distribution
(\ref{vdistr}) of volumes $v=V_X/V_X^{\rm max}$ when all $a_i = a =
7/3n$:
\begin{eqnarray*}
 d\CN[v]/dv &=& \sum_{\tilde{N_i}\geq 1} \delta(v - \prod_i \tilde{N}_i^{-a})\\
 &\approx& \int_{\tilde{N}_i \geq 1} d^n \tilde{N} \, \delta(v - \prod_i \tilde{N}_i^{-a}) \\
 &=& \int_{U_i \geq 0} d^n U  \, e^{\sum_j U_j} \, \delta(v-e^{-a \sum_j
 U_j}) \\
 &=& \frac{1}{a}\, e^{-(1+a) \frac{\ln v}{a}} \int d^n U \, \delta(\sum_j U_j + \frac{\ln
 v}{a})\\
 &=& \frac{1}{a}\, v^{-(1+1/a)} \, \frac{(-\frac{\ln
 v}{a})^{n-1}}{(n-1)!} \, \Theta(1-v)\\
 &=& \frac{\left(\frac{3n}{7}\right)^n}{(n-1)!} (-\ln
 v)^{n-1} \, v^{-\frac{3n}{7}-1} \, \Theta(1-v)
\end{eqnarray*}

\end{document}